\documentstyle[12pt,epsf]{article}

\newcommand{\beq}{\begin{equation}}
\newcommand{\eeq}{\end{equation}} 
\newcommand{\beqa}{\begin{eqnarray}}
\newcommand{\eeqa}{\end{eqnarray}}
\newcommand{\nn}{\nonumber}

\def\ifmath#1{\relax\ifmmode#1\else$#1$\fi}
\def\to{\rightarrow}

\def\BR{\mbox{{\rm BR}}}
\def\Re{\mbox{{\rm Re}}}
\def\Im{\mbox{{\rm Im}}}
\def\arg{{\rm Arg}}
\def\gsim{{~\raise.15em\hbox{$>$}\kern-.85em
          \lower.35em\hbox{$\sim$}~}}
\def\lsim{{~\raise.15em\hbox{$<$}\kern-.85em
          \lower.35em\hbox{$\sim$}~}}
\def\B{{\cal B}}
\def\adir{a_{\rm dir}}
\def\eff{_{\rm eff}}

\begin{document}


\title{\bf Taming the Penguin in the $B^0_d(t)\to \pi^+\pi^-$ CP-Asymmetry: Observables and Minimal Theoretical Input}
\author{\textsc{J\'er\^ome Charles}}
\date{June 1998 (revised September 1998)}
\maketitle

\begin{center}
Laboratoire de Physique Th\'eorique et Hautes \'Energies~\footnote{Laboratoire associ\'e au Centre National de la Recherche
Scientifique - URA D00063\\
E-mail: charles@qcd.th.u-psud.fr}
\\
Universit\'e de Paris-Sud, B\^atiment 210, F-91405 Orsay Cedex, France
\end{center}

\thispagestyle{empty}\setcounter{page}{0}

\vskip 5 mm
\begin{flushright}
LPTHE-Orsay 98-35\\
hep-ph/9806468
\end{flushright}

\begin{abstract}
Penguin contributions, being not negligible in general, can hide
the information on the
CKM angle $\alpha$ coming from the measurement of the time-dependent $B^0_d(t)\to\pi^+\pi^-$ CP-asymmetry. Nevertheless, we show that this information can be summarized in a set of simple equations, expressing $\alpha$ as a multi-valued function of a single theoretically unknown
parameter, which conveniently can be chosen as a well-defined ratio of penguin to tree amplitudes. Using these exact analytic expressions, free of any assumption besides the Standard Model, and some reasonable hypotheses to constrain the modulus of the penguin amplitude,
we derive several new upper bounds on the penguin-induced shift $|2\alpha-2\alpha\eff|$, 
generalizing the recent result of Grossman and Quinn. These bounds depend on the
average branching ratios of some decays ($\pi^0\pi^0$, $K^0\overline{K^0}$, $K^\pm\pi^\mp$) particularly sensitive to the penguin. On the other hand, with further and less conservative approximations,
we show that the knowledge of the $B^\pm\to K\pi^\pm$ branching ratio alone gives
sufficient information to extract the free parameter without the need of other
measurements, and without knowing $|V_{td}|$ or $|V_{ub}|$. More generally, knowing the modulus of the penguin amplitude
with an accuracy of $\sim 30\%$ might result in an extraction of $\alpha$ competitive
with the experimentally more difficult isospin analysis. We also show that our 
framework allows to recover most of the previous approaches in a transparent and simple
way, and in some cases to improve them. In addition we discuss in detail the problem
of the various kinds of discrete ambiguities.
\end{abstract}

\newpage

\section{Introduction}
In a close future, several collaborations---\textsc{BaBar}, BELLE, CDF, CLEO, HERA-B---will hopefully
make the first measurements of CP-violation in the $B_d$ system~\cite{burasRev}.
The most important consequences concerning the Standard Model (SM) would be the determination of the Unitarity Triangle (UT). However, if the measurement of the UT angle $\sin2\beta$ seems to be straightforward
from both experimental and theoretical points of view thanks to the very clean
$B\to J/\Psi K_S$ decay, the extraction of $\alpha$ from the standard mode $B\to\pi^+\pi^-$
is still an open problem~\footnote{Throughout this paper, $B$ stands for a $B_d$ meson.}. Since it has been pointed out that QCD---and mixed
QCD/electroweak---radiative corrections (called ``penguins'')
induce potentially large theoretical uncertainties on this angle~\cite{1st}, many papers
have been devoted to this subject~\cite{others}.

In a pioneering paper~\cite{GL}, Gronau and London have shown that the knowledge
of the $B(\overline B)\to\pi^+\pi^-,\pi^0\pi^0,\pi^\pm\pi^0$ branching ratios leads to
the determination of the gluonic penguin effects, assuming isospin symmetry and neglecting electroweak penguin contributions.
Then, with this information and the usual mixing-induced CP-asymmetry it is possible to get $\alpha$ up
to discrete ambiguities. The main drawback of this interesting method is the expected
smallness of the $B\to\pi^0\pi^0$ branching ratio ($10^{-7}$--$10^{-6}$) due to
colour-suppression. This fact, combined with the detection efficiency of the final state
and the needed tagging of the flavour of the $B$-meson, constitutes a difficult challenge
to $e^+e^-$ $B$-factories and an almost impossible task for future hadronic machines---LHCb, BTeV.

Then it was realized by Silva and Wolfenstein~\cite{silvWolf} that by extending
the flavour symmetry to SU(3) one can gain further information on penguin effects,
the key point being the $K\pi$ modes where the ratio penguin/tree is certainly greater than 1.
Considering the crudeness of the assumptions made in the original paper in addition to SU(3), the method has been extended until a high level of sophistication by
several authors~\cite{gronauSU3}. As a consequence, it is not clear to what
extent such complicated geometrical constructions, plagued by multiple discrete
ambiguities, are sensitive to $\alpha$ and to
the unavoidable theoretical assumptions. Therefore these strategies will give
conservative results only when a better understanding of non-leptonic $B$-decays is
available. In addition, two simpler SU(3) approaches concerning $\alpha$ have been
proposed by Buras and Fleischer~\cite{burasK0K0b} and Fleischer and Mannel~\cite{FMzoology} respectively, which will be discussed
in more detail below.

One can also use a model---usually factorization---to estimate the penguin amplitude,
and then compute the difference between $\alpha$ at the input and $\alpha\eff$ at the output, as Aleksan et al.~\cite{ABLOPR}
and Ciuchini et al.~\cite{rome1} did, or directly get a model-dependent $\alpha$ as was proposed
by Marrocchesi and Paver~\cite{paver}~\footnote{Actually we will see that the Marrocchesi-Paver method~\cite{paver} is
essentially the same as the Fleischer-Mannel~\cite{FMzoology} one, although the
theoretical input is different.}.

Thus, after having {\it hunted}~\cite{burasHunt}, {\it trapped}~\cite{LNQS} and made the 
{\it zoology}~\cite{FMzoology} of the penguin, it is time to begin {\it taming} it. To
accomplish this task, we first remark that most of the authors cited above have
computed {\it the observables}---branching ratio and CP-asymmetries---{\it as functions of the
theoretical parameters}---QCD matrix elements and CKM factors, including the
angle $\alpha$. We will
follow {\it the opposite way}, and show that it is indeed a fruitful approach. Although fully
equivalent to the ``traditional'' one, it leads to a very important and simple
new result: {\it it is possible to express independently of any
model~\footnote{In this paper, ``model-independent'' means ``not relying on a particular
hadronic model which describes non-perturbative physics''. On the contrary, we
will assume that the SM holds for the parametrization of CP-asymmetries and amplitudes.},
and in an exact and simple way all the theoretical parameters,
including the angle $\alpha$, as functions of the experimentally accessible observables and
of only one real theoretical unknown.} The latter can be chosen as, e.g., $|P/T|$, the ratio of
``penguin'' to ``tree'' amplitudes (which are {\it unambiguously} defined below).
It is also possible to use as the unknown a pure QCD quantity, free of any dependence
with respect to $|V_{td}|$ or $|V_{ub}^\ast|$ contrary to the parameter $|P/T|$; in the latter case, we give polynomial
equations directly expressed in the $(\rho,\eta)$ plane.
We have exploited
these exact analytic expressions to derive several new and simple results and to recover
some of the previous approaches. The main points of this paper are:
\begin{itemize}
\item
Using the exact parametrization in terms of $|P/T|$, it is possible to represent the information
given by the time-dependent CP-asymmetry in the $(|P/T|,2\alpha)$ plane. Of course
without any further assumption on the magnitude of $|P/T|$ there is no way to
constrain $\alpha$. But this $(|P/T|,2\alpha)$ plot provides a nice transparent presentation of experimental data, where our ignorance of the strong interactions
is relegated to a single parameter.
\item
As soon as one is interested in quantifying the size of the penguin---and indeed we
are, {\it $\sin2\alpha$ is not a good parameter.} One should simply use $2\alpha$ instead.
Actually using $\sin2\alpha$ rather than $2\alpha$ is not wrong, but {\it one loses half
of the information} as we will see in detail below.
This is already true at the level of the parametrization in terms of $|P/T|$, and {\it this is also
true for all the methods allowing to remove the penguin effects}, which give generically $2\alpha$ rather
than $\sin2\alpha$, up to discrete ambiguities. To make clear this point which up to now has remained
confused, we will treat explicitly the example of the Gronau-London isospin analysis. On the contrary, the observables
depend only on $2\alpha$ or equivalently on $\tan\alpha$, and thus the $\alpha\to
\pi+\alpha$ ambiguity is always present~\cite{quinnDisAmb}.
\item
Bounding the magnitude of the penguin allows directly to bound the shift of the
CKM angle $\alpha$ from the directly observable $\alpha\eff$. This can be done using
information from decays particularly sensitive to the penguin. For example, assuming SU(2) isospin
symmetry and neglecting electroweak penguin contributions we are able to derive two bounds depending on $\BR(B\to\pi^0\pi^0)$, one of which being the Grossman-Quinn bound~\cite{GQbound} while the other is new. Assuming the larger SU(3) symmetry,
we obtain two new bounds depending on $\BR(B\to K^0\overline{K^0})$ and $\lambda^2\BR(B\to K^\pm\pi^\mp)$ respectively which, not surprisingly, may be more constraining than the
SU(2) ones, and which need some, {\it but not all}, the usual assumptions concerning the neglect of annihilation
and/or electroweak penguin diagrams. As far as the branching ratios of the penguin-sensitive
modes are concerned, these bounds do not need flavour tagging and are still valid when only
an upper limit on the branching ratios is available. In addition, they can be slightly modified to be used when the actual value of the direct CP-asymmetry in the $B\to\pi^+\pi^-$ channel is not available,
as it is shown below.
Depending on the actual values of the branching
ratios, the theoretical error on $\alpha$ constrained by these bounds could be as large
as $\sim 30^\circ$ or as small as $\sim 10^\circ$. In particular, the most recent CLEO
analyses of the $\pi^+\pi^-$ and $K^\pm\pi^\mp$ modes~\cite{CLEO} allow us to give
for the first time the following numerical bound
\beq
|2\alpha-2\alpha\eff|\le\Delta\,,\ \ \ \mbox{with}\ \ \ 25^\circ<\Delta<59^\circ
\eeq
assuming rather weak hypotheses in the SU(3) limit (see \S~\ref{K+pi-Section}) and 
BR$(B\to\pi^+\pi^-)>0.4\times 10^{-5}$ in addition to the CLEO data.
\item
Finally, after having stressed that only one hadronic parameter has to be 
estimated by the theory in order to get $\alpha$, we give one new explicit example: 
assuming SU(3) and neglecting annihilation and electroweak penguin diagrams, we
show that $\BR(B^\pm\to K\pi^\pm)$
gives sufficient information to solve a degree-four polynomial equation in the
$(\rho,\eta)$ plane, which roots can be represented as curves in this
plane. Contrary to the Fleischer-Mannel proposal~\cite{FMzoology}, ours does not need the knowledge of $|V_{td}|$ or $|V_{ub}|$, and
requires only the measurement of $\BR(B^\pm\to K\pi^\pm)$ in addition to the usual
time-dependent $B\to\pi^+\pi$ time-dependent CP-asymmetry. Alternatively, the
knowledge of the modulus of the penguin amplitude (or the ratio of penguin to tree)
with an uncertainty of $\sim 30\%$ should provide a rather good estimation of
$\alpha$. This kind of strategy,
although affected by potentially large theoretical uncertainties, may be necessary
when the more conservative bounds are too weak to be really useful in testing the SM.
\end{itemize}

The paper is organized as follows: in Section~\ref{summary}, we summarize the main results of this work---this section should be of immediate use
for the reader not interested
by the development. In Section~\ref{basics} we fix our notations
in writing the general parametrization of the amplitudes. With the help of the recent CLEO 
measurements of non-leptonic charmless $B$-decays, we give some rough orders of magnitude
of the expected penguin pollution. Then we derive the
equations giving the theoretical parameters, including $\alpha$, as functions of the
observables and the theoretical unknown, treated first as a free parameter,
and latter eventually constrained under reasonable hypotheses. For example in Section~\ref{pi0pi0Section} we show how
to use in our framework the information coming from the $B\to\pi^0\pi^0$ and $B^\pm\to\pi^\pm\pi^0$
decays, to obtain the
Grossman-Quinn bound and a new similar isospin bound. In Section~\ref{SU3section}
we exhibit two new bounds, based on the SU(3) assumption, which may be 
more stringent than the two isospin bounds. Then in Section~\ref{K0pi+Section} we discuss
an explicit example where the theoretical unknown is actually estimated rather than
bounded. A reasonable knowledge of $\alpha$ can be expected even if one allows a
sizeable violation of the theoretical assumptions. In Section~\ref{previous}
we discuss how to incorporate and improve some of the previous approaches in our language, and clarify some points which have been mistreated in the literature,
in particular the problem of the discrete ambiguities. Our conclusion is that although the penguin-induced error on $\alpha$ is expected to be quite large in the
$B\to\pi^+\pi^-$ channel, {\it it should be under the control of the theory}. Therefore the
generalization of the methods presented here to other channels is very desirable
to get more constraints on $\alpha$.

This paper has two technical appendices: the first one~(\ref{model}) explains how we
got the values of the observables from a naive calculation, in order to numerically illustrate
our purpose before experimental data is available and the second
one~(\ref{2aeffbar}), following Grossman and Quinn~\cite{GQbound}, shows explicitly the existence of bounds which are independent
of the measurement of the direct CP-asymmetry.
\section{Summary}
\label{summary}
\subsection{Exact Model-Independent Results}
Defining the Standard Model $B\to\pi^+\pi^-$ amplitudes
\beqa
A(B^0\to\pi^+\pi^-)&=&V_{ud}V_{ub}^\ast M^{(u)}+V_{td}V_{tb}^\ast M^{(t)}
\ =\ e^{+i\gamma}T+e^{-i\beta}P \,, \label{def1}\\
A(\overline{B^0}\to\pi^+\pi^-)&=&V_{ud}^\ast V_{ub} M^{(u)}+V_{td}^\ast V_{tb} M^{(t)}
\ =\ e^{-i\gamma}T+e^{+i\beta}P \,,
\eeqa
the time-dependent $B^0(t)\to\pi^+\pi^-$ CP-asymmetry
\beq\label{aCP}
a_{\rm CP}(t)=\adir\cos\Delta m\,t-\sqrt{1-\adir^2}\sin2\alpha\eff \sin\Delta m\,t \,,
\eeq
and the average $B,\overline{B}\to f,\bar f$ branching ratio
\beq
\B_{f,\bar f}=\frac{1}{2}\left[\BR(B\to f)+\BR(\overline{B}\to\bar f)\right]\,,
\eeq
we prove in the following that the Standard Model predicts very simple relations between $\alpha$ and
$|P/T|$, $|P|$, $|T|$ and $\delta=\arg(PT^\ast)$ respectively, these relations depending
only on the observables $\B_{\pi^+\pi^-}$, $\adir$ and $2\alpha\eff$ and being completely
free of any assumption on hadronic physics:
\beqa
\cos(2\alpha-2\alpha\eff)&=&\frac{1}{\sqrt{1-\adir^2}}\left [ 1-\left ( 1-\sqrt{1-\adir^2}\cos2\alpha\eff\right )\left | \frac{P}{T} \right |^2\, \right ]\,,\label{master1}\\
|P|^2&=&\frac{\B_{\pi^+\pi^-}}{1-\cos2\alpha}\left [ 1-\sqrt{1-\adir^2}\cos(2\alpha-2\alpha\eff)\right ]\,,\label{master2}\\
|T|^2&=&\frac{\B_{\pi^+\pi^-}}{1-\cos2\alpha}\left [ 1-\sqrt{1-\adir^2}\cos2\alpha\eff\right ]\,,\label{master3}\\
\tan\delta&=&\frac{\adir\tan\alpha}{1-\sqrt{1-\adir^2}\left [ \cos2\alpha\eff+\tan\alpha\sin2\alpha\eff\right ]}\,.\label{master4}
\eeqa

Rather than $|P/T|$, $|P|$ and $|T|$ which incorporate respectively a $|V_{td}/V_{ub}^\ast|$, $|V_{td}|$ and $|V_{ub}^\ast|$ factor, one may prefer to
write Eqs.~(\ref{master1}-\ref{master3}) in terms of $|M^{(t)}/M^{(u)}|$,
$|M^{(t)}|$ and $|M^{(u)}|$ respectively (see the definition~(\ref{def1})).
As $|V_{td}|$ and $|V_{ub}^\ast|$ depend also on the UT, it is not possible to
express such relations as functions of $\alpha$ alone; instead we use the
Wolfenstein parametrization and find three polynomial equations in the
$(\rho,\eta)$ plane. With the definitions of the following combinations of observables
\beq
D_c\equiv \sqrt{1-\adir^2}\cos2\alpha\eff\,,\ \ \ \ \ 
D_s\equiv \sqrt{1-\adir^2}\sin2\alpha\eff\,,
\eeq
and the theoretical parameters $|M^{(t)}|$ and $|M^{(u)}|$ normalized to
$\sqrt{\B_{\pi^+\pi^-}}/|\lambda V_{cb}|$
\beq
\frac{R_P}{R_T}=\left|\frac{M^{(t)}}{M^{(u)}}\right|^2\,,\ \ \ \ \ 
R_P\equiv |\lambda V_{cb}|^2\frac{|M^{(t)}|^2}{\B_{\pi^+\pi^-}}\,,\ \ \ \ \ 
R_T\equiv |\lambda V_{cb}|^2\frac{|M^{(u)}|^2}{\B_{\pi^+\pi^-}}\,,
\eeq
one has two degree-four polynoms depending respectively on $R_P/R_T$ and $R_P$
\beqa
&&(1-D_c)(1-\frac{R_P}{R_T})\,\rho^4+2(1-D_c)(1-\frac{R_P}{R_T})\,\rho^2\eta^2+(1-D_c)(1-\frac{R_P}{R_T})\,\eta^4 \nn\\
&&-2(1-D_c)(1-2\frac{R_P}{R_T})\,\rho^3-2D_s\,\rho^2\eta-2(1-D_c)(1-2\frac{R_P}{R_T})\,\rho\eta^2-2D_s\,\eta^3 \nn\\
&&+(1-D_c)(1-6\frac{R_P}{R_T})\,\rho^2+2D_s\,\rho\eta+[1+D_c-2(1-D_c)\frac{R_P}{R_T}]\,\eta^2\nn\\
&&+(1-D_c)\frac{R_P}{R_T}\,(4\rho-1)=0 \label{master5}\,,
\eeqa
\beqa
&&(1-D_c)\,\rho^4+2(1-D_c-R_P)\,\rho^2\eta^2+(1-D_c-2R_P)\,\eta^4 \nn\\
&&-2(1-D_c)\,\rho^3-2D_s\,\rho^2\eta-2(1-D_c-2R_P)\,\rho\eta^2-2D_s\,\eta^3 \nn\\
&&+(1-D_c)\,\rho^2+2D_s\,\rho\eta+(1+D_c-2R_P)\,\eta^2=0 \label{master6}\,,
\eeqa
and one linear equation depending on $R_T$ (the $\pm$ sign being related to a discrete ambiguity) 
\beq
\sqrt{1-D_c}\,(\rho-1) \pm \sqrt{2R_T-1+D_c}\,\,\eta=0 \label{master7}\,.
\eeq

Eqs.~(\ref{master5}-\ref{master7}) are another way of writing Eqs.~(\ref{master1}-\ref{master3}) by replacing respectively $|P/T|$, $|P|$ and $|T|$
by the ratios $R_P/R_T$, $R_P$ and $R_T$: the advantage is that the latter parameters
do not depend on the badly known CKM matrix elements $|V_{td}|$ and $|V_{ub}^\ast|$.
\subsection{Phenomenological Applications}
\label{summaryPheno}
It has become standard in the CP-literature to use several phenomenological
assumptions, some of which can be very good while some others can be strongly
violated. As a result, it is often not easy for the reader to know exactly which
approximations are used by the authors, and thus to make his own opinion about the
accuracy of these theoretical prejudices. In this paper, we will try to state clearly
what kind of hypotheses we use in addition to the SM; some of the results that we
derive rely on a few reasonable assumptions chosen in the list below.

\begin{itemize}
\item {\bf Assumption 1}
$|P/T|< 1$. This very conservative bound should be distinguished from the small penguin expansion.\label{PsurTinf1.a}
\item {\bf Assumption 2}
SU(2) isospin symmetry of the strong interactions.\label{SU2.a}
\item {\bf Assumption 3}
SU(3) flavour symmetry of the strong interactions.\label{SU3.a}
\item {\bf Assumption 4}
Neglect of the OZI-suppressed annihilation penguin diagrams (cf. Fig.~\ref{fig:penguins}).\label{zweig.a}
\item {\bf Assumption 5}
Neglect of the electroweak penguin contributions.\label{Pew.a}
\item {\bf Assumption 6}
Neglect of the $V_{us}V_{ub}^\ast$ contributions to the $B^+\to K^0\pi^+$ amplitude.\label{TK0Pi+.a}
\end{itemize}

\paragraph{Upper Bounds.} We have found several quantities bounding the shift of the
true $2\alpha$ from the experimentally accessible $2\alpha\eff$, among which~(\ref{GQbound2}) is the 
Grossman-Quinn bound~\cite{GQbound}, while the others are new:

\beqa\label{bound1}
\mbox{if }\sin2\alpha\eff > 0 && 0 \,<\, 2\alpha \,<\, 2\pi-2\arcsin(\sin2\alpha\eff) \nn\\
\mbox{if }\sin2\alpha\eff < 0 && -2\arcsin(\sin2\alpha\eff) \,<\, 2\alpha \,<\, 2\pi
\ \ \ \ \ \mbox{[assuming~{\bf  1}],}
\eeqa

\beq\label{GQbound2}
|2\alpha-2\alpha\eff| \le \arccos \left [ \frac{1}{\sqrt{1-\adir^2}}
\left ( 1-2\frac{\B_{\pi^0\pi^0}}{\B_{\pi^\pm\pi^0}} \right )\right ]
\ \ \ \ \ \mbox{[assuming~{\bf  2} and~{\bf  5}],}
\eeq

\beq
|2\alpha-2\alpha\eff| \le \arccos \left [ \frac{1}{\sqrt{1-\adir^2}}
\left ( 1-4\frac{\B_{\pi^0\pi^0}}{\B_{\pi^+\pi^-}} \right )\right ]
\ \ \ \ \ \mbox{[assuming~{\bf  2} and~{\bf  5}],}
\eeq

\beq
|2\alpha-2\alpha\eff| \le \arccos \left [ \frac{1}{\sqrt{1-\adir^2}}
\left ( 1-2\frac{\B_{K^0\overline{K^0}}}{\B_{\pi^+\pi^-}} \right )\right ]
\ \ \ \ \ \mbox{[assuming~{\bf  3} and~{\bf  5}],}
\eeq

\beq\label{bound5}
|2\alpha-2\alpha\eff| \le \arccos \left [ \frac{1}{\sqrt{1-\adir^2}}
\left ( 1-2\lambda^2\,\frac{\B_{K^\pm\pi^\mp}}{\B_{\pi^+\pi^-}} \right )\right ]
\ \ \ \ \ \mbox{[assuming~{\bf  3} and~{\bf  4}].}
\eeq

Following Grossman and Quinn~\cite{GQbound} we show that, under the same hypotheses, the above upper bounds still hold if one replaces in Eqs.~(\ref{bound1}-\ref{bound5}) $\adir$ by zero and 
$2\alpha\eff$ by $2\overline{\alpha}\eff$ where the latter effective angle is defined by
\beq\label{a2effBar}
{\rm sign}(\cos2\overline{\alpha}\eff)\equiv{\rm sign}(\cos2\alpha)\,,\ \ \ \ \ 
\sin2\overline{\alpha}\eff\equiv\sqrt{1-\adir^2}\sin2\alpha\eff\,.
\eeq
For example, one has the bound
\beq
|2\alpha-2\overline{\alpha}\eff| \le \arccos  \left ( 1-2\lambda^2\,\frac{\B_{K^\pm\pi^\mp}}{\B_{\pi^+\pi^-}} \right )
\ \ \ \ \ \mbox{[assuming~{\bf  3} and~{\bf  4}],}
\eeq
and so on.
Although these bounds on $|2\alpha-2\overline{\alpha}\eff|$ are weaker than the ones
on $|2\alpha-2\alpha\eff|$, they have the advantage that they do not depend on
the measurement of $\adir$: indeed, the angle $2\overline{\alpha}\eff$ is accessible (up to a twofold ambiguity) from the 
$\sin\Delta m\,t$ term only (see Eqs.~(\ref{aCP}) and~(\ref{a2effBar})). Therefore the experimental
uncertainty should be smaller for $2\overline{\alpha}\eff$ than for $2\alpha\eff$.

In addition to these upper bounds, we derive a {\it lower} bound on $|2\alpha-2\alpha\eff|$
in \S~\ref{lowerBound}, to which we refer the reader for more details.

\paragraph{Determination of $\alpha$.} We propose a new method for the
extraction of $\alpha$---up to discrete ambiguities, which improves the Fleischer-Mannel proposal~\cite{FMzoology}.

The idea (often used in the literature) is to estimate the modulus of the penguin
contribution with the help of the $B^\pm\to K\pi^\pm$ decay. We avoid the problem
of knowing $|V_{td}|$ by using directly the polynomial equation~(\ref{master6}) in the
$(\rho,\eta)$ plane, with the theoretical parameter $R_P$ given by
\beq
R_P=\lambda^2\frac{\B_{K\pi^\pm}}{\B_{\pi^+\pi^-}}
\ \ \ \ \ 
\mbox{[assuming~{\bf  3},~{\bf  4},~{\bf  5} and~{\bf  6}].}
\eeq
This typically leads to draw four allowed curves in the $(\rho,\eta)$ plane, which
in the limit $R_P\to 0$, reduce to the two circles representing the no-penguin solution
$\sin2\alpha=\sin2\alpha\eff$.
\section{Theoretical Framework}
\label{basics}
\subsection{Standard Model Parametrization of the Amplitudes}
\label{notation}
The aim of this paragraph is to recall some already known results and to fix the notation used in this paper.

The time-dependent rate for an oscillating state $B^0(t)$ which has been tagged as a $B^0$ 
meson
at time $t=0$ is given by (for simplicity the $e^{-\Gamma t }$ and constant phase space factors
are omitted below~\footnote{Tiny differences between phase space of the various channels discussed in this paper are neglected.})
\beq\label{timeRate1}
\Gamma\left ( B^0(t)\to\pi^+\pi^- \right ) = \frac{|A|^2+|\bar A|^2}{2}+\frac{|A|^2-|\bar A|^2}{2}
\cos\Delta m\,t-\Im\left (\frac{q}{p}\bar AA^\ast\right )\sin\Delta m\,t \,,
\eeq
where
\beq
A\equiv A(B^0\to\pi^+\pi^-)\,,\ \ \ \ \ \bar A\equiv A(\overline{B^0}\to\pi^+\pi^-)
\eeq
and $q/p=\exp(-2i\beta)$ in the Wolfenstein phase convention, which
provides an expansion of the CKM matrix in powers of $\lambda\equiv |V_{us}|\sim 0.22$~\cite{wolf}.
With this convention, one has
\beq
\beta=\arg(-V_{td}^\ast)\,,\ \ \ \ \ \gamma=\arg(-V_{ub}^\ast)\,,\ \ \ \ \ \arg(V_{ts})={\cal O}(\lambda^2)\,,
\eeq
while the other CKM matrix elements are real (up to highly-suppressed $\lambda^n$ terms) and the angle $\alpha$ is given by
$\alpha=\pi-\beta-\gamma$. Defining
\beqa
\B_{\pi^+\pi^-}&\equiv&\frac{1}{2}\left [ \BR(B^0\to\pi^+\pi^-)+\BR(\overline{B^0}\to\pi^+\pi^-)\right ]\,,\label{B+-}\\
\adir&\equiv&\frac{|A|^2-|\bar A|^2}{|A|^2+|\bar A|^2}\,,\label{adir}\\
2\alpha\eff&\equiv&\arg\left ( \frac{q}{p}\bar AA^\ast\right )\,,\label{2aeff}
\eeqa
the rate~(\ref{timeRate1}) becomes
\beq\label{timeRate2}
\Gamma\left ( B^0(t)\to\pi^+\pi^- \right )=\B_{\pi^+\pi^-}\left [ 1+\adir \cos\Delta m\,t
-\sqrt{1-\adir ^2}\sin2\alpha\eff \sin\Delta m\,t\right ]\,.
\eeq
The time-dependent CP-asymmetry reads
\beqa
a_{\rm CP}(t)&\equiv&\frac{\Gamma\left ( B^0(t)\to\pi^+\pi^- \right )-\Gamma\left ( \overline{B^0}(t)\to\pi^+\pi^- \right )}{\Gamma\left ( B^0(t)\to\pi^+\pi^- \right )+\Gamma\left ( \overline{B^0}(t)\to\pi^+\pi^- \right )}\nn\\
&=&\adir\cos\Delta m\,t-\sqrt{1-\adir^2}\sin2\alpha\eff \sin\Delta m\,t \,.\label{timeCP}
\eeqa
We may define another effective angle by~\footnote{As the sign of $\cos2\overline{\alpha}\eff$ is not observable, it can be defined arbitrarily. However,
the exact definition is important for the derivation of the bounds (cf. Appendix~\ref{2aeffbar}).}
\beq\label{2aeffbarDef}
{\rm sign}(\cos2\overline{\alpha}\eff)\equiv{\rm sign}(\cos2\alpha)\,,\ \ \ \ \ 
\sin2\overline{\alpha}\eff\equiv\sqrt{1-\adir^2}\sin2\alpha\eff
\eeq
such as
\beq\label{timeCP2}
a_{\rm CP}(t)=\adir\cos\Delta m\,t-\sin2\overline{\alpha}\eff\sin\Delta m\,t\,.
\eeq
$\adir$ is the direct CP-asymmetry, while $\sin2\overline{\alpha}\eff=\sqrt{1-\adir^2}\sin2\alpha\eff$ is the
mixing-induced CP-asymmetry. In the absence of penguins, one has $\adir=0$ and
$\sin2\overline{\alpha}\eff=\sin2\alpha\eff=\sin2\alpha$. The experiment allows the measurement of three fully
model-independent observables, one is CP-invariant ($\B_{\pi^+\pi^-}$), while the two others
are CP-asymmetries ($\adir$ and $\sin2\alpha\eff$, or $\adir$ and $\sin2\overline{\alpha}\eff$).

It has often been assumed in the literature that the dominant penguin amplitude
is the top-mediated one, with the consequence that this amplitude
is proportional to $V_{td}V_{tb}^\ast$. This assumption
has received a lot of attention recently~\cite{fleischerPuPc,rome}. In any case, 
the $u$-penguin contributions which are proportional to $V_{ud}V_{ub}^\ast$ (as well as other contributions such as exchange diagrams)
can be incorporated in the definition of the ``tree'' amplitude. Contributions proportional to
$V_{cd}V_{cb}^\ast$---``charming penguins''~\cite{rome}---can be rewritten, by CKM unitarity ($V_{cd}V_{cb}^\ast=-V_{ud}V_{ub}^\ast-V_{td}V_{tb}^\ast$), in terms of the two other
combinations. Thus, just from the weak phase structure of the SM, we may write the $B^0\to\pi^+\pi^-$ physical amplitude as
\beqa
A&=&V_{ud}V_{ub}^\ast M^{(u)}+V_{td}V_{tb}^\ast M^{(t)}\nn\\
&\equiv&e^{+i\gamma}T_{\pi^+\pi^-}+e^{-i\beta}P_{\pi^+\pi^-}\label{TP}
\eeqa
and similarly for the other $2\pi$ channels~\footnote{Note that $P_{\pi^+\pi^0}$ comes from electroweak penguins, and/or from isospin symmetry breaking.}
\beqa
A_{\pi^0\pi^0}\equiv A(B^0\to\pi^0\pi^0)&\equiv&\frac{1}{\sqrt{2}}(e^{+i\gamma}\,T_{\pi^0\pi^0}+e^{-i\beta}P_{\pi^0\pi^0})\,,\label{TPpi0pi0}\\
A_{\pi^+\pi^0}\equiv A(B^+\to\pi^+\pi^0)&\equiv&\frac{1}{\sqrt{2}}(e^{+i\gamma}\,T_{\pi^+\pi^0}+e^{-i\beta}P_{\pi^+\pi^0})\label{TPpi+pi0}\,.
\eeqa
Let us stress that there is {\it absolutely no approximation} in writing
Eq.~(\ref{TP}-\ref{TPpi+pi0}): $T_{\pi\pi}$ and $P_{\pi\pi}$ are CP-conserving complex quantities, defined by the
weak phase that they carry, and they
incorporate all possible SM topologies such as trees, penguins, electroweak
penguins... In this sense, many (not all, however) of the methods proposed previously for the extraction of $\alpha$ in the top-dominance assumption are in fact still
valid if the latter hypothesis is relaxed~\footnote{Of course, numerical estimates of quantities like $|P/T|$ may be
greatly modified by charming penguins~\cite{fleischerPuPc,rome}.}. The CP-conjugate channels are obtained by reversing the sign of the weak phases:
\beq\label{TPbar}
\bar A_{\pi\pi}=e^{-i\gamma}T_{\pi\pi}+e^{+i\beta}P_{\pi\pi}\,.
\eeq
As said above, neglecting penguin diagrams ($P_{\pi\pi}=0$) gives
\beq
2\alpha\eff=\arg\left ( \frac{q}{p}T_{\pi^+\pi^-}T_{\pi^+\pi^-}^\ast e^{-2i\gamma}\right )=2\alpha\,.
\eeq
From now on we will denote in the whole paper
\beq\label{TP2}
T_{\pi^+\pi^-}\equiv T\,,\ \ \ \ \ P_{\pi^+\pi^-}\equiv P\,,
\eeq
and {\it by economy of language, we will call the $T$ and $P$ amplitudes ``tree'' and
``penguin'' respectively,} although $T$ gets contributions from $u$- and $c$-penguin
diagrams~\footnote{Be careful that our definition of ``tree'' and ``penguin'' amplitudes, relying on CP-phases, is slightly different from the one used in Refs.~\cite{burasHunt,fleischerNew}, although the consequence is the same: these so-defined amplitudes
are unambiguous and physical quantities.}.

In this paper, we will also consider the $B\to K^0\overline{K^0}$, $B\to K^\pm\pi^\mp$ and $B^\pm\to K\pi^\pm$ decays. 
For the former, we adopt the following parametrization
\beqa
A(B^0\to K^0\overline{K^0})&=&V_{ud}V_{ub}^\ast M_{K\overline{K}}^{(u)}+V_{td}V_{tb}^\ast M_{K\overline{K}}^{(t)} \nn\\
&\equiv&e^{+i\gamma}\,T_{K\overline{K}}+e^{-i\beta}P_{K\overline{K}}\,,\label{TPK0K0b}
\eeqa
while for the latter it is convenient to expand on the CKM basis
$(V_{us}V_{ub}^\ast,V_{cs}V_{cb}^\ast)$ (recall that $V_{cs}V_{cb}^\ast$ is real in the Wolfenstein convention):
\beqa
A(B^0\to K^+\pi^-)&=&V_{us}V_{ub}^\ast M_{K^+\pi^-}^{(u)}+V_{cs}V_{cb}^\ast M_{K^+\pi^-}^{(c)} \nn\\
&\equiv&e^{+i\gamma}\,T_{K^+\pi^-}+P_{K^+\pi^-}\,,\label{TPK+pi-}\\
A(B^+\to K^0\pi^+)
&\equiv&e^{+i\gamma}\,T_{K^0\pi^+}+P_{K^0\pi^+}\,.\label{TPK0pi+}
\eeqa
As far as the $B\to K^0\overline{K^0}$ amplitude is concerned, we have used the notation $T_{K\overline{K}}$ to make apparent
the resemblance with the other channels; however it should be stressed that this decay is a pure penguin process. Actually $T_{K\overline{K}}$ represents the
contribution of the long-distance $u$- and $c$-penguins~\cite{fleischerK0K0b}.

Let us repeat that Eqs.~(\ref{TP}-\ref{TPbar}) and~(\ref{TPK0K0b}-\ref{TPK0pi+}) rely only on the Standard Model.

\subsection{General Bounds}
Similarly to Eq.(~\ref{B+-}), we will denote by $\B_{f,\bar f}$ the CP-conserving average branching
ratio
\beq\label{BR}
\B_{f,\bar f}=\frac{1}{2}\left[\BR(B\to f)+\BR(\overline B\to\bar f)\right]\,.
\eeq
For example,
\beq
\B_{K\pi^\pm}=\frac{1}{2}\left[\BR(B^+\to K^0\pi^+)+\BR(B^-\to\overline{K^0}\pi^-)\right]
\eeq
and so on.

From the discussion in \S~\ref{notation}, it is clear that the SM predicts each $B^0\to f$ decay
amplitude as the sum of two terms carrying two different CP-violating phases
$\phi_1$, $\phi_2$:
\beq
A(B^0\to f)=e^{+i\phi_1}M_1+e^{+i\phi_2}M_2\,,\ \ \ \ \ A(\overline{B^0}\to \bar{f})=\eta_f\left(e^{-i\phi_1}M_1+e^{-i\phi_2}M_2\right)
\eeq
where $M_1$ and $M_2$, although complex numbers, are CP-conserving and the sign $\eta_f$ depends
on the CP of the final state, e.g. $\eta_f(\pi^+\pi^-)=+$. Thus the
average branching ratio~(\ref{BR}) writes
\beq
\B_{f,\bar f}=|M_1|^2+|M_2|^2+2\,\Re(M_1M_2^\ast)\cos(\phi_1-\phi_2)\,.
\eeq
Note that we express the amplitudes squared in ``units of two-body branching ratio''. For fixed values of $M_2$ and $\phi_1-\phi_2$, $\B_{f,\bar f}$ as a function of
$M_1$ takes its minimal value when $M_1=-M_2\cos(\phi_1-\phi_2)$, in which case
$\B_{f,\bar f}|_{\rm min}=|M_2|^2\sin^2(\phi_1-\phi_2)$. Interverting the r\^oles
of $M_1$ and $M_2$, we obtain the following (exact) general bounds:
\beq\label{galBounds}
|M_1|^2\sin^2(\phi_1-\phi_2)\,\le\,\B_{f,\bar f}\,,\ \ \ \ \ \ 
|M_2|^2\sin^2(\phi_1-\phi_2)\,\le\,\B_{f,\bar f}\,.
\eeq
Such inequalities have been previously employed for the demonstration of the so-called
Fleischer-Mannel bound~\cite{FMbound}. We will use Eq.~(\ref{galBounds}) extensively
throughout this paper.
\subsection{Orders of Magnitude}
\label{magnitude}
The recently updated CLEO analyses of $B$-decays into two light pseudoscalars give precious information on the
various quantities discussed in this paper~\cite{CLEO}:
\beqa
\B_{\pi^+\pi^-}&<&0.84\times 10^{-5} \ \ [90\%\mbox{ CL}]\,,\nn\\
\B_{K^\pm\pi^\mp}&=&(1.4\pm0.3\pm0.2)\times 10^{-5}\,,\label{CLEO}\\
\B_{K\pi^\pm}&=&(1.4\pm0.5\pm0.2)\times 10^{-5}\,.\nn
\eeqa
Thanks to this experimental information, it is possible to derive a crude lower bound
for the ratio $|P/T|$. Indeed, from Eqs.~(\ref{TP}) and~(\ref{galBounds}) we have
\beq
|T|^2\sin^2\alpha\,\le\,\B_{\pi^+\pi^-}\,.
\eeq
Furthermore, while the $\pi\pi$ penguin is proportional to $V_{td}V_{tb}^\ast$, 
the $K\pi$ penguin is proportional to $V_{cs}V_{cb}^\ast$ (cf. Eqs.~(\ref{TP})
and~(\ref{TPK+pi-}-\ref{TPK0pi+})). Thus we have
\beq
|P|\sim\left|\frac{V_{td}}{V_{cb}^\ast}\right|\times \sqrt{\B_s}
\eeq
where $\B_s$ is a typical scale of the $B\to K\pi$ branching ratios, assuming that
these channels are dominated by the QCD penguin, and that the QCD-part of the penguin matrix
elements are of the same order for $\pi\pi$ and $K\pi$. Thus we obtain
\beq
\left|\frac{P}{T}\right|\,\gsim\,\lambda|\sin\gamma|\sqrt{\frac{\B_s}{\,\B_{\pi^+\pi^-}}}
\eeq
where we have used $|V_{td}/V_{cb}|=\lambda\sin\gamma/\sin\alpha$.

Numerically, from the current SM constraints on the $(\rho,\eta)$ parameters~\cite{LAL}, we have $|\sin\gamma|\gsim 0.6$. The CLEO data~(\ref{CLEO}) suggest
$\B_s/\B_{\pi^+\pi^-}\gsim 1/0.84$ which gives
\beq
\left|\frac{P}{T}\right|\,\gsim\,0.14\,.
\eeq
Note that contrary to some claims, factorization gives typically $|P/T|\sim 0.15$~\cite{ABLOPR} and is not ruled out by the CLEO data yet, although it is only
marginally compatible.

Although this calculation is only illustrative, it is clear that the penguin 
contributions pose a serious problem for the extraction of $\alpha$ from the CP-asymmetry. More complete analyses show that the ratio $|P/T|$ can easily be
$\sim$ 30\% or even 40\%~\cite{rome}. Therefore one can not avoid to define theoretical
procedures allowing to reduce the penguin uncertainty, or at least to control it.
This is the subject of the present paper.

Let us add here one comment on the size of the direct CP-asymmetry. With a ratio $|P/T|$ of
order ${\cal O}(20$--$30\%)$, it is straightforward to show that a strong phase
$\arg(PT^\ast)$ of order ${\cal O}(10$--$20^\circ)$ is sufficient to generate a direct CP-asymmetry of order ${\cal O}(10\%)$ (cf. Appendix~\ref{model}). While perturbative calculations of such phases predict
in general very small values~\cite{BSS}, it is likely that non-perturbative effects will
considerably enhance these Final State Interactions (FSI)~\footnote{Note also that in the $N_c\to\infty$ limit, since $u$- and $c$-penguin can contribute, such phases between $T$ and $P$
amplitudes are in principle ${\cal O}(1)$. Indeed as a perturbative calculation suggests~\cite{fleischerPuPc}, the real and imaginary parts of the long-distance penguins are of the same order.}~\cite{donoghue}. Thus we expect that $\sin2\alpha\eff$ and $\adir$
will be measured with comparable statistical accuracy~\cite{babarBook}.

At this stage, we refer the reader to Appendix~\ref{model}, where we calculate
the relevant parameters and observables in a naive way, in order to numerically
illustrate the phenomenological results that we derive below.
\subsection{Some Exact One-Parameter Results}
\label{masterSection}
Let us first consider only the $B^0(t)\to\pi^+\pi^-$ rate, Eq.~(\ref{timeRate2}). As pointed out above,
there are three observables: the average branching
ratio and the
CP-asymmetries 
\beq\label{observables}
\B_{\pi^+\pi^-}\,,\ \ \ \ \ \adir\ \ \ \ \ \mbox{and}\ \ \ \ \ \sin2\alpha\eff\,.
\eeq
From $\sin2\alpha\eff$, one gets $2\alpha\eff$ up to
a twofold discrete ambiguity. While the vanishing of the penguin amplitude $P$
implies $2\alpha\eff=2\alpha$, the SM description of the rate~(\ref{timeRate2}) involves four parameters, three of which are CP-conserving while the
fourth is the CP-violating angle $\alpha$ (see Eqs.~(\ref{TP}) and~(\ref{TPbar})), namely
\beq\label{4param}
|T|\,,\ \ \ \ \ |P|\,,\ \ \ \ \ \delta=\arg(PT^\ast)\,,\ \ \ \ \ \mbox{and}\ \ \ \ \ \alpha\,,
\eeq
one overall phase being irrelevant, and after the use of $q/p=\exp(-2i\beta)$.
Thus the presence of the penguin amplitude forbids the measurement of $\alpha$, the
number of parameters being greater than the number of observables. However, as we
will see, it is possible to express $\alpha$ in terms of the three observables
and of one of the four parameters. The latter can be chosen as either $|P/T|$, $|P|$, $|T|$, or $\delta$.

From~(\ref{TP}) and~(\ref{TPbar}), we deduce
\beqa
-(2i\sin\alpha) P&=&e^{-i\gamma}A-e^{i\gamma}\bar A\,,\\
(2i\sin\alpha) T&=&e^{i\beta}A-e^{-i\beta}\bar A\,,
\eeqa
that can be rewritten as
\beqa
(2i\sin\alpha) P&=&e^{i\beta}\left [ e^{i\alpha}A-e^{-i\alpha}\frac{q}{p}\bar A \right ]\,,\label{masterP}\\
(2i\sin\alpha) T&=&e^{i\beta}\left [ A-\frac{q}{p}\bar A \right ]\,.\label{masterT}
\eeqa
Calculating the ratio $|P/T|^2$ from~(\ref{masterP}-\ref{masterT}) and using the
definitions~(\ref{adir}-\ref{2aeff}) we obtain the very important, although
very simple, equation:
\beq\label{masterEq1}
\cos(2\alpha-2\alpha\eff)=\frac{1}{\sqrt{1-\adir^2}}\left [ 1-\left ( 1-\sqrt{1-\adir^2}\cos2\alpha\eff\right )\left | \frac{P}{T} \right |^2\, \right ]\,.
\eeq
Thus {\it Eq.~(\ref{masterEq1}) defines $2\alpha$ as a four-valued function of $|P/T|$:} indeed, $2\alpha\eff$ is known
up to a twofold discrete ambiguity, while Eq.~(\ref{masterEq1}) also contains a
twofold discrete ambiguity as long as $2\alpha$ is concerned, because of the cosine function. Note that these two
discrete ambiguities are of a different nature: the $2\alpha\eff\to\pi-2\alpha\eff$ ambiguity
is inherent to CP-eigenstate analyses, while the $2\alpha-2\alpha\eff\to-(2\alpha-2\alpha\eff)$
ambiguity is generated by the penguin contributions. One can also view the latter
ambiguity by saying that the no-penguin solution $2\alpha=2\alpha\eff$ appears as a double root of the general cosine equation~(\ref{masterEq1}), which degeneracy is lifted by the penguin contributions. 

Let us add here one comment on the discrete ambiguity generated by getting $\alpha$
from $2\alpha$, that is the $\alpha\to \pi+\alpha$ ambiguity. From~(\ref{B+-}-\ref{2aeff}),~(\ref{TP}), (\ref{TPbar}) and~(\ref{4param}) one sees that the three observables $\B_{\pi^+\pi^-}$, $\adir$ and $2\alpha\eff$ are invariant under the transformation
$\alpha\to \pi+\alpha,\,\delta\to \pi+\delta$. Thus these observables depend only on
$2\alpha$ and $2\delta$, or equivalently on $\tan\alpha$ and $\tan\delta$. Without
any further assumption on the strong phase, the $\alpha\to \pi+\alpha$ ambiguity is
irreducible~\cite{quinnDisAmb}, the signs of $\sin\alpha$ and $\sin\delta$ being
related by the equation:
\beq
{\rm sign}(\sin\alpha)\times{\rm sign}(\sin\delta)={\rm sign}(\adir)\,.
\eeq
As far as the SM is accepted, the latter ambiguity not a real problem because the constraints on the UT select only one of the two ambiguous solutions---we already know that $0<\alpha<\pi$ because $\eta$ is positive~\cite{LAL}---and moreover because these solutions cannot merge as they are separated
by $\pi$. This is not the case, obviously, for the ambiguity $2\alpha\eff\to\pi-2\alpha\eff$. In the following we will always express our results in terms
of $2\alpha$ and $2\delta$, or $\tan\alpha$ and $\tan\delta$.

It is also possible to derive very simple relations expressing the parameters
$|P|$, $|T|$ and $\tan\delta$ as functions of $|P/T|$, or equivalently, as functions
of $2\alpha$ which is itself a function of $|P/T|$ through Eq.~(\ref{masterEq1}). Indeed, from~(\ref{B+-}-\ref{2aeff}),~(\ref{4param}) and~(\ref{masterP}-\ref{masterT}) we get~\footnote{The fact that only the weak angle $\alpha$ is present in Eqs.~(\ref{masterEq1}) and~(\ref{masterEq2}-\ref{masterEq4}) originates from the peculiar SM prediction
that the $B^0-\overline{B^0}$ mixing is dominated by the top-loop, just cancelling the
CP-phase of the penguin defined by Eq.~(\ref{TP}), and is not related to the dominance (or not)
of the top in penguin loops. This is obviously not the case,
e.g., for the decay $B\to K_S\pi^0$ where both $\alpha$ and $\beta$ enter in the game.}:
\beqa
|P|^2&=&\frac{\B_{\pi^+\pi^-}}{1-\cos2\alpha}\left [ 1-\sqrt{1-\adir^2}\cos(2\alpha-2\alpha\eff)\right ]\,,\label{masterEq2}\\
|T|^2&=&\frac{\B_{\pi^+\pi^-}}{1-\cos2\alpha}\left [ 1-\sqrt{1-\adir^2}\cos2\alpha\eff\right ]\,,\label{masterEq3}\\
\tan\delta&=&\frac{\adir\tan\alpha}{1-\sqrt{1-\adir^2}\left [ \cos2\alpha\eff+\tan\alpha\sin2\alpha\eff\right ]}\,.\label{masterEq4}
\eeqa
Note the following important point:
Eq.~(\ref{masterEq2}) (resp. Eq.~(\ref{masterEq3}))
gives $2\alpha$ as a four-valued function of $|P|$ (resp. $|T|$), and
Eq.~(\ref{masterEq4}) gives $2\alpha$ as a bi-valued function of $2\delta$; this
feature puts the parameters $|P/T|$, $|P|$, $|T|$ and $2\delta$ on an equal footing:
each of these is candidate to be the single theoretical input.

Let us discuss the no-penguin limit of Eqs.~(\ref{masterEq1}) and~(\ref{masterEq2}-\ref{masterEq4}):
$|P|\to 0$ (and thus $\adir\to 0$). In this limit, Eqs.~(\ref{masterEq1}) and~(\ref{masterEq2}) reduce
simply to $2\alpha=2\alpha\eff$, as it should. Eq.~(\ref{masterEq3}) reduce to
$|T|^2=\B_{\pi^+\pi^-}$ independently of $\alpha$, as also expected. Finally
Eq.~(\ref{masterEq4}) becomes indefinite in this limit, because $\delta$ itself
becomes indefinite. Note the important point that the parameters $|P/T|$ and $|P|$
measure directly the size of the penguin and thus of the shift $|2\alpha-2\alpha\eff|$ while the parameters $|T|$ and $\delta$ carry only poor
information on the size of the penguin (for example the no-penguin relation
$|T|^2=\B_{\pi^+\pi^-}$ can still be verified with a non-vanishing $|P|$, for particular values of the parameters).

Lastly, we stress that Eqs.~(\ref{masterEq1}-\ref{masterEq4}) are fully
{\it equivalent} to the original Eqs.~(\ref{TP}) and~(\ref{TPbar}) together with the
definitions~(\ref{B+-}-\ref{2aeff}) and~(\ref{4param}). They are exact relations 
between the theoretical parameters taken two by two, depending only on the 
observables. The most obvious application of this result is the modelization
of the theoretical error induced by penguin effects on the CKM angle. For
example the reader may choose his favourite model or his favourite assumptions to estimate $|P/T|$ as well as its
associated error. This range of values of $|P/T|$ propagates into four cleanly-defined, although model-dependent, discrete
solutions for $2\alpha$ and their theoretical errors, thanks to Eq.~(\ref{masterEq1}). The same can be done using, instead $|P/T|$ and Eq.~(\ref{masterEq1}), the parameters $|P|$ and Eq.~(\ref{masterEq2}), or $|T|$ and
Eq.~(\ref{masterEq3}), or $2\delta$ and
Eq.~(\ref{masterEq4})~\footnote{Note that the extraction of $\alpha$ from Eq.~(\ref{masterEq4}) would require a very accurate and unlikely knowledge of
$\delta$.}.
As the model is used to calculate {\it only one} real
quantity, this procedure should be safer than the ones proposed in,
e.g., Refs.~\cite{ABLOPR} and ~\cite{rome1}. 
We will give practical examples of this strategy in Sections~\ref{K0pi+Section} and~\ref{previous}.

Another application, which is more conservative but less informative, is to use
channels where the penguin may be dominant and, when related by a flavour symmetry
to the parameter $|P|$, can help to bound the shift $|2\alpha-2\alpha\eff|$.
This is the case for $B\to\pi^0\pi^0$, $K^0\overline{K^0}$ and $K^\pm\pi^\mp$, as
explained in Sections~\ref{pi0pi0Section} and~\ref{SU3section}. But before that,
we shall insist now on the mo\-del-in\-de\-pen\-dent features of Eqs.~(\ref{masterEq1}) and~(\ref{masterEq2}-\ref{masterEq4}).
\subsection{Plotting the CP-Asymmetry in the $\left(\left|\frac{P}{T}\right|,2\alpha\right)$ Plane}
From $|\cos(2\alpha-2\alpha\eff)|\le 1$ and from~(\ref{masterEq1}) the following allowed
interval for $|P/T|^2$ is obtained:
\beq\label{PsurTintervalle}
\frac{1-\sqrt{1-\adir^2}}{1-\sqrt{1-\adir^2}\cos2\alpha\eff}
\le \left | \frac{P}{T} \right |^2 \le
\frac{1+\sqrt{1-\adir^2}}{1-\sqrt{1-\adir^2}\cos2\alpha\eff}\,.
\eeq
The lower bound on $|P/T|^2$ is induced by the direct CP-asymmetry---it becomes trivial in the limit $\adir\to 0$; indeed, if the
latter is non zero, then a non-vanishing penguin amplitude follows. As the sign
of $\cos2\alpha\eff$ is not observable, (\ref{PsurTintervalle}) defines actually
two different intervals for $|P/T|$, one for the branch corresponding to $\cos2\alpha\eff>0$ and the other for the branch corresponding to $\cos2\alpha\eff<0$.

\begin{figure}[ht]
\centerline{\epsfxsize=0.8\textwidth\epsfbox{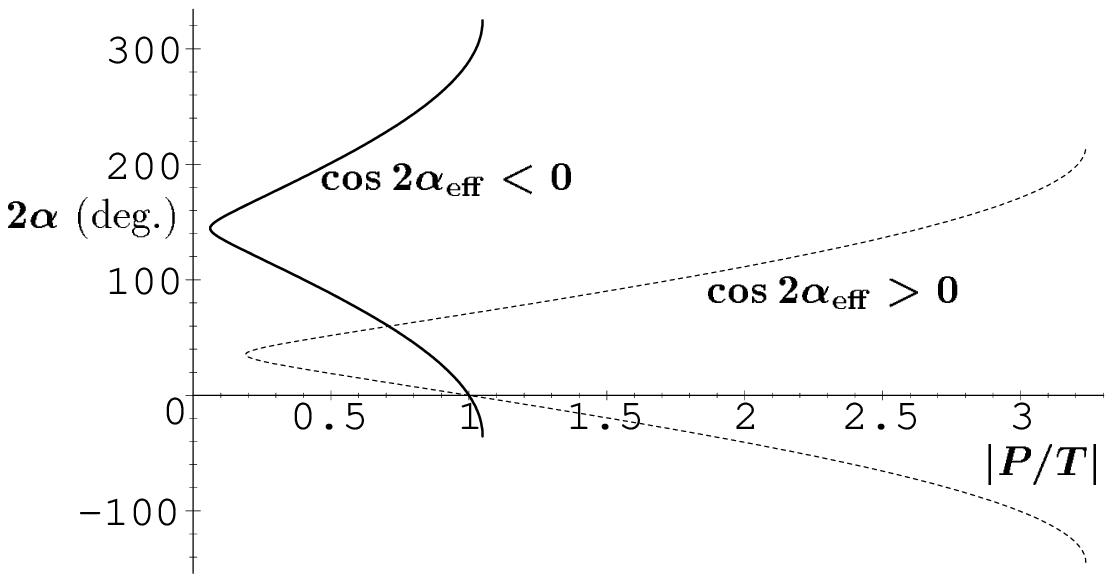}}
\caption{\it The CKM angle $2\alpha$ as a function of $|P/T|$ obtained from Eq.~(\ref{masterEq1}), for $\adir=0.12$ and $\sin2\alpha\eff=0.58$ (see Appendix~\ref{model} for a description of this numerical example). The solid curve corresponds to $\cos2\alpha\eff<0$ and
the dashed one to $\cos2\alpha\eff>0$. Note that $|P/T|$ varies in the range~(\ref{PsurTintervalle}) which depends on the sign of $\cos2\alpha\eff$.
For a given value of $|P/T|$, there are in 
general four solutions for $2\alpha$.}
\label{fig:alpha}
\end{figure}

Assuming $\adir$ and $\sin2\alpha\eff$ have been measured, Eq.~(\ref{masterEq1}) allow to plot $2\alpha$ as a function of $|P/T|$ varying
in the interval~(\ref{PsurTintervalle}): on Fig.~\ref{fig:alpha} we represent the two distinct
branches corresponding to the two possible signs for $\cos2\alpha\eff$.

\begin{figure}[ht]
\centerline{\epsfxsize=0.49\textwidth\epsfbox{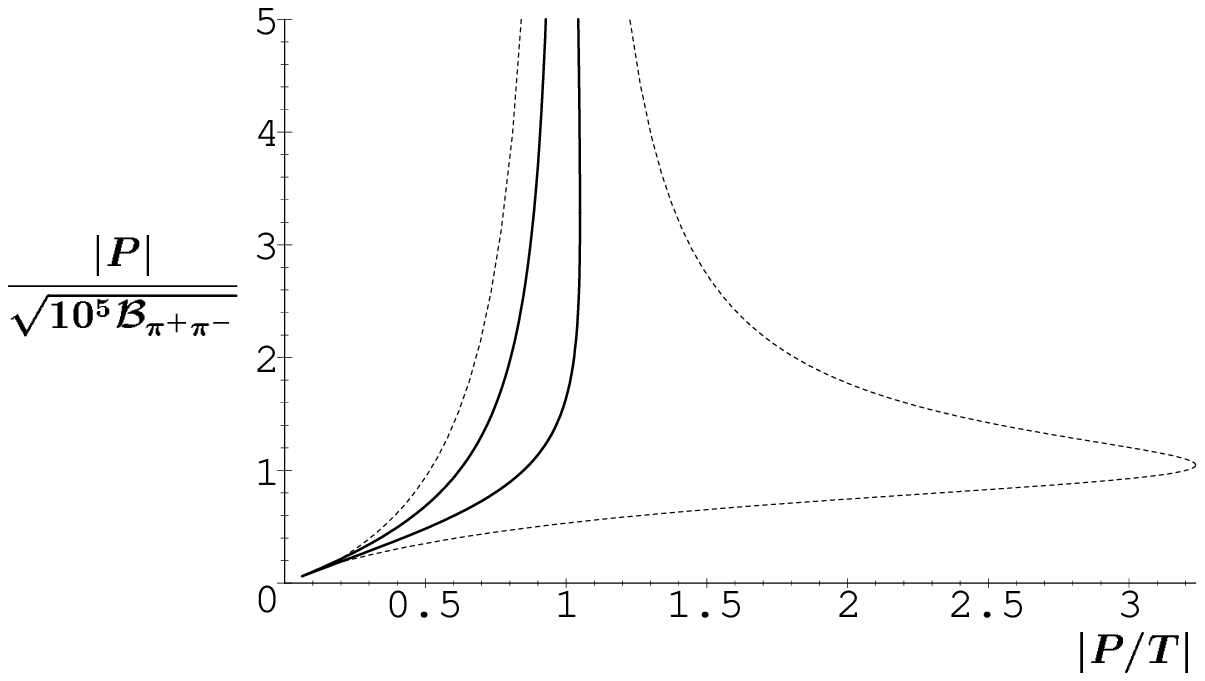}\hspace{0.01\textwidth}
\epsfxsize=0.49\textwidth\epsfbox{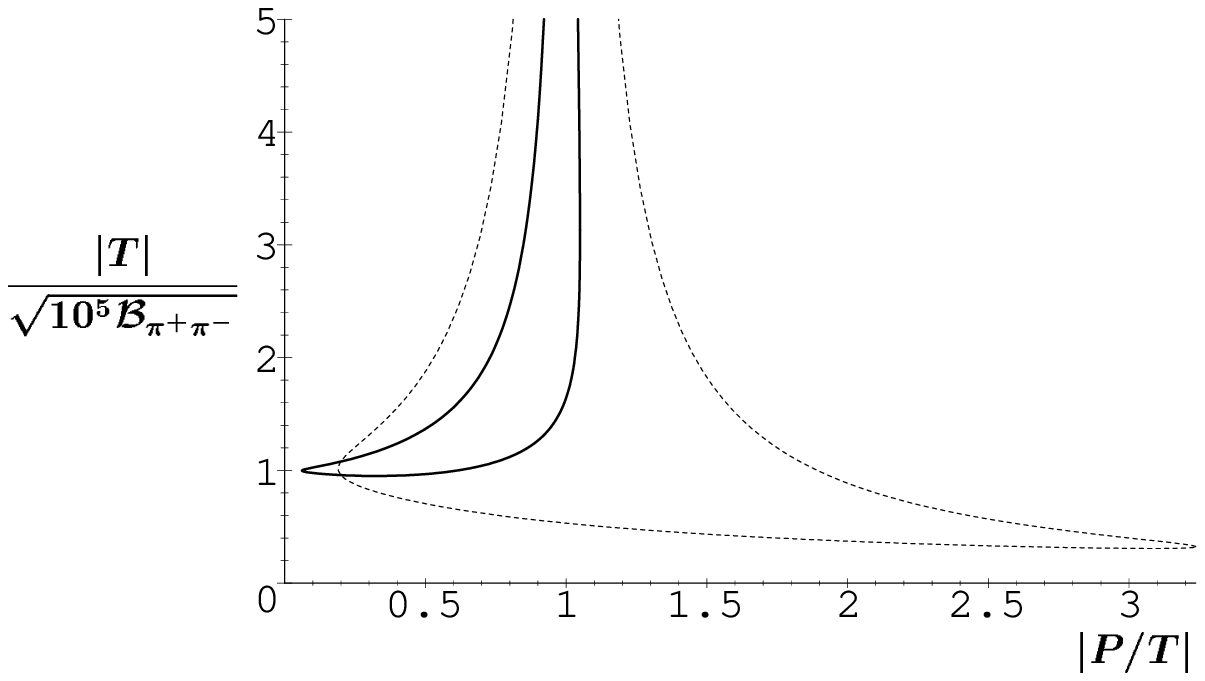}}
\centerline{\epsfxsize=0.49\textwidth\epsfbox{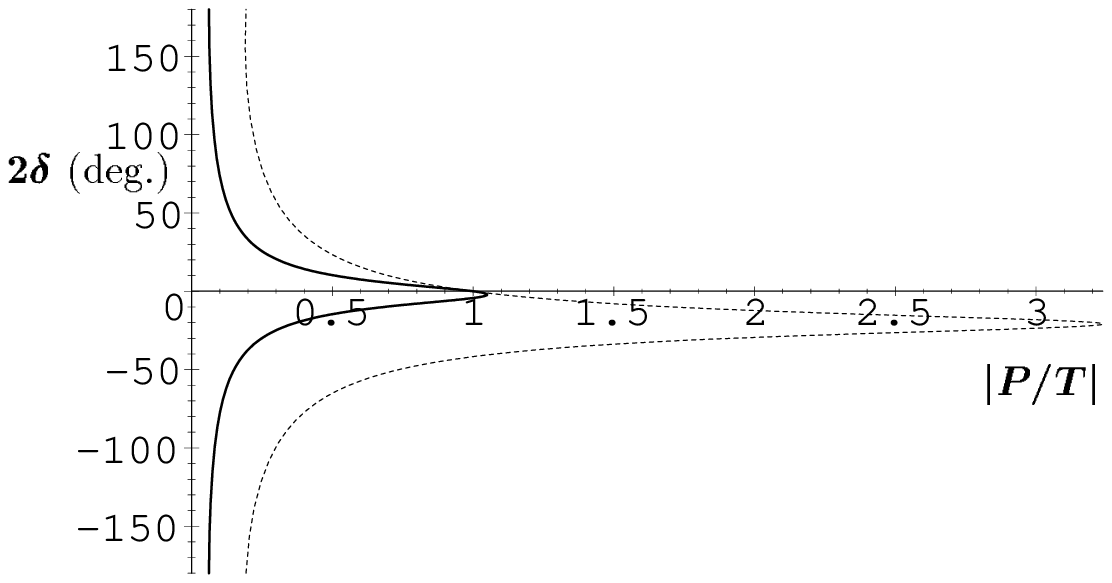}}
\caption{\it The $|P|$, $|T|$ amplitudes and the
strong phase $2\delta$ respectively as functions of $|P/T|$ obtained from Eqs.~(\ref{masterEq1}) and~(\ref{masterEq2}-\ref{masterEq4}), for the numerical example $\adir=0.12$ and $\sin2\alpha\eff=0.58$ (see Appendix~\ref{model}). The solid curves correspond to $\cos2\alpha\eff<0$ and
the dashed ones to $\cos2\alpha\eff>0$. Note that $|P|$ and $|T|$ diverge when
$2\alpha\to 0$, $|P/T|\to 1$ because the CP-asymmetries $\adir$ and $\sin2\alpha\eff$
are kept fixed: $|P|$ and $|T|$ have to be very large to produce CP-violation with a
small CP-phase $\alpha$.}
\label{fig:PTdelta}
\end{figure}

Furthermore, the three remaining equations~(\ref{masterEq2}-\ref{masterEq4}) together
with $2\alpha$ given by Fig.~\ref{fig:alpha}
allow to represent $|P|/\sqrt{\B_{\pi^+\pi^-}}$, $|T|/\sqrt{\B_{\pi^+\pi^-}}$, and $2\delta$ as functions of $|P/T|$ (Fig.~\ref{fig:PTdelta}).

Let us summarize the main properties and virtues of Eqs.~(\ref{masterEq1}) and~(\ref{masterEq2}-\ref{masterEq4}) and of Figs.~\ref{fig:alpha}-\ref{fig:PTdelta}:
\begin{itemize}
\item
These are absolutely {\it exact} results, relying only on the SM, and provide
a nice representation of what kind of model-independent information can be obtained
from the measurement of the time-dependent $B^0(t)\to\pi^+\pi^-$ CP-asymmetry {\it only}.
The $(|P/T|,2\alpha)$ plot may be used to present the experimental results which
hopefully will be available from $B$-factories in the next years. 
\item
Fig.~\ref{fig:alpha} shows that the linear approximation in $|P/T|$, which is used in some
papers~\cite{gronauPsurT,FMzoology} is indeed very good for $|P/T|\lsim 1$ as far as 
$2\alpha-2\alpha\eff$ is concerned. However,
Eqs.~(\ref{masterEq1}) and ~(\ref{masterEq2}-\ref{masterEq4}) expanded to first order in $|P/T|$ give
expressions which are not particularly simpler, and thus it is more convenient to keep the
exact formul\ae.
\item
Eqs.~(\ref{masterEq1}) and~(\ref{masterEq2}-\ref{masterEq4}) are not invariant under the 
transformation $\alpha\to\frac{\pi}{2}-\alpha$. {\it This does not properly mean that the $\alpha\to\frac{\pi}{2}-\alpha$ ambiguity is lifted}: the souvenir of this ambiguity lives in the invariance of Eqs.~(\ref{masterEq1}) and~(\ref{masterEq2}-\ref{masterEq4}) under
$\alpha_{\rm eff}\to\frac{\pi}{2}-\alpha_{\rm eff}$ because {\it the sign of $\cos2\alpha\eff$ is
not known.} {\it This means however that $\sin2\alpha$ is not a good parameter}: indeed the
penguin effect is not the same for the solutions corresponding to $\cos2\alpha\eff>0$
than for the others corresponding to $\cos2\alpha\eff<0$, as Fig.~\ref{fig:alpha} clearly
shows. In particular, the solutions corresponding to $\cos2\alpha\eff<0$ are more
affected by the penguin uncertainty which is an important information~\footnote{This has already been noticed by Gronau in Ref.~\cite{gronauPsurT}.}.
As long as
the penguin is not strong enough to change the sign of $\cos2\alpha\eff/\cos2\alpha$, we
actually expect $\cos2\alpha\eff<0$ from the current SM constraints on the UT~\cite{LAL}. To be illustrative, let us plot $\sin2\alpha$ as a function of
$|P/T|$ using Eq.~(\ref{masterEq1}) and compare with $2\alpha$ as a function of
$|P/T|$ (Fig.~\ref{fig:sin2a}).
Obviously, four curves in the $(|P/T|,\sin2\alpha)$ plane gives half as less
information than four curves in the $(|P/T|,2\alpha)$ plane. If we reconstruct the
$|P/T|\to2\alpha$ curves from the $|P/T|\to\sin2\alpha$ ones, we will get {\it eight} curves
among which four are ``wrong'' solutions (Fig.~\ref{fig:sin2a}). We discuss further this point in \S~\ref{GronauLondon},
with the explicit example of the Gronau-London construction.
We conclude that {\it one should not express the penguin effect in terms
of $\sin2\alpha-\sin2\alpha\eff$} as it is sometimes done in the literature~\cite{rome1,paver}.
\begin{figure}[ht]
\centerline{\epsfxsize=0.49\textwidth\epsfbox{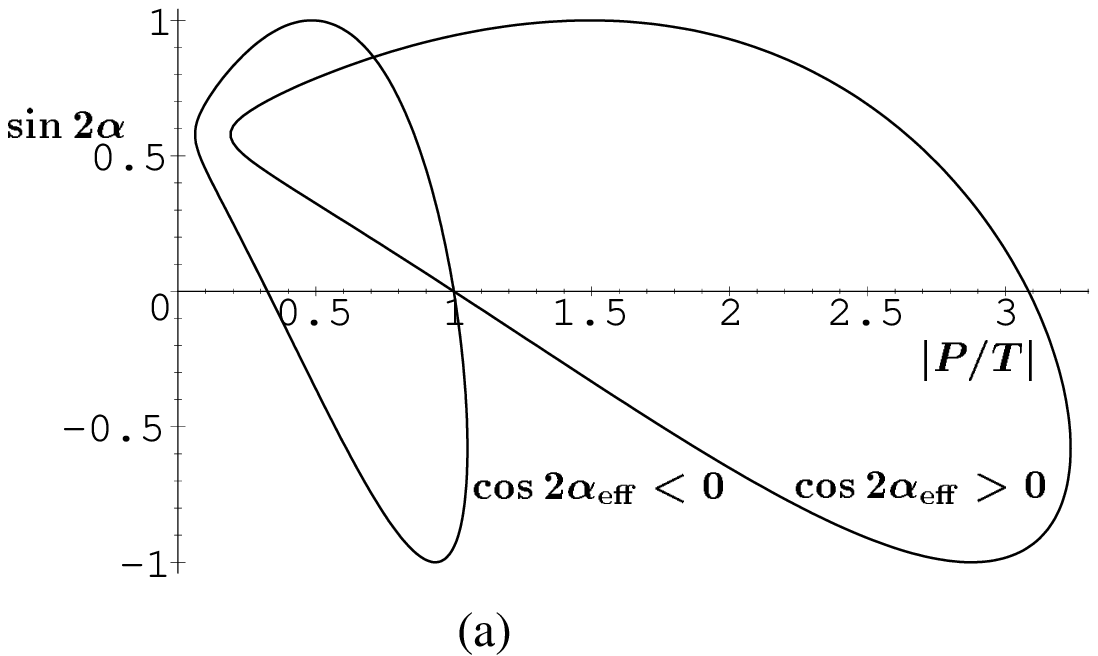}\hspace{0.01\textwidth}
\epsfxsize=0.49\textwidth\epsfbox{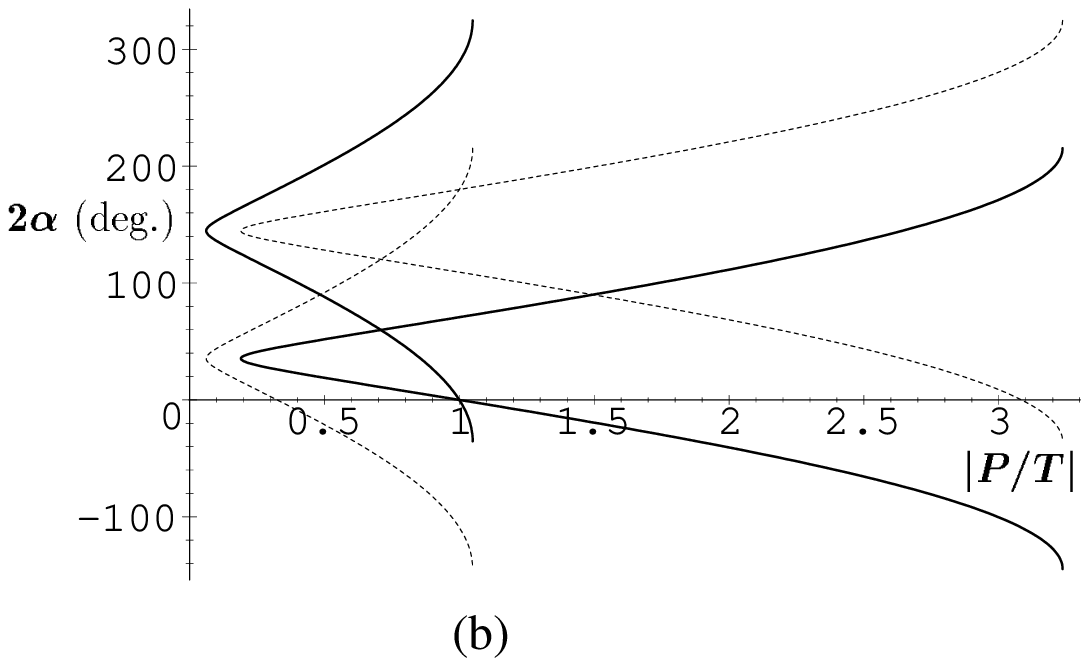}}
\caption{\it {\rm(a)} $\sin2\alpha$ as a function of $|P/T|$ obtained from Eq.~(\ref{masterEq1}), for the numerical example $\adir=0.12$ and $\sin2\alpha\eff=0.58$ (see Appendix~\ref{model}). The two branches corresponding to the two possible signs for
$\cos2\alpha\eff$ are not degenerate.  {\rm(b)} $2\alpha$ as a function of $|P/T|$ obtained by computing $\arcsin(\sin2\alpha)$ and
$\pi-\arcsin(\sin2\alpha)$: the comparison with Fig.~\ref{fig:alpha} shows that
the dashed curves are wrong solutions.}
\label{fig:sin2a}
\end{figure}

\item
Bounding the absolute magnitude of the penguin directly allows to bound the shift
$|2\alpha-2\alpha\eff|$ (and {\it vice-versa}) thanks to Eq.~(\ref{masterEq1}) or
Eq.~(\ref{masterEq2}). For
example, the very conservative estimate $|P/T|< 1$ (Assumption~{\bf  1}) leads to the simple bound
\beq
\cos(2\alpha-2\alpha\eff) > \cos2\alpha\eff\,.
\eeq
Of course, this bound does not allow a precise measurement of $\alpha$. 
Nevertheless, with only a very weak assumption, it provides
an allowed interval for $2\alpha$ (we have taken into account that the sign of
$\cos2\alpha\eff$ is not known):
\beqa
\mbox{if }\sin2\alpha\eff > 0 && 0 \,<\, 2\alpha \,<\, 2\pi-2\arcsin(\sin2\alpha\eff) \nn\\
\mbox{if }\sin2\alpha\eff < 0 && -2\arcsin(\sin2\alpha\eff) \,<\, 2\alpha \,<\, 2\pi
\ \ \ \ \ \mbox{[assuming~{\bf  1}].\label{1ereBorne}}
\eeqa
As explained in Appendix~\ref{2aeffbar}, this bound implies a weaker one, obtained by replacing above $2\alpha\eff$ by $2\overline{\alpha}\eff$ where the latter effective angle is defined by Eq.~(\ref{2aeffbarDef}).
The only advantage in using $2\overline{\alpha}\eff$ instead of $2\alpha\eff$  is that
the former angle follows directly (up to a twofold ambiguity) from the $\sin\Delta m\,t$ term in the time-dependent CP-asymmetry~(\ref{timeCP2}) independently of 
$\adir$; thus the experimental uncertainty on $2\overline{\alpha}\eff$ is expected
to be smaller than on $2\alpha\eff$.

For the numerical example that we have chosen (Appendix~\ref{model}), $\sin2\alpha\eff=0.58$, we obtain from the bound~(\ref{1ereBorne})
$0<2\alpha<290^\circ$, which is of course not very informative.
\end{itemize}
\subsection{Exact One-Parameter Polynoms in the $(\rho,\eta)$ Plane}
\label{RhoEtaPlane}
For evaluation by hadronic models, or by using phenomenological assumptions such as
in Section~\ref{K0pi+Section}, the theoretical parameters $|P/T|$, $|P|$ and $|T|$ may be
not suitable as they depend on QCD matrix elements times $|V_{td}/V_{ub}^\ast|$, $|V_{td}|$ and $|V_{ub}^\ast|$ 
respectively. Indeed the latter CP-conserving CKM factors are badly known, therefore
they would introduce an additional uncertainty in combination with the theoretical model-induced
error for the estimation of the QCD-part of the matrix elements. In the literature, this problem
has been solved by scanning the whole allowed domain for $(\rho,\eta)$~\cite{ABLOPR},
by simply assuming that such CKM factors would be known from other measurements~\cite{FMzoology}, or by expressing $|V_{td}/V_{ub}^\ast|$ as a function
of $\alpha$ and $\beta$, the latter angle being determined from future CP-measurements
in the $B\to J/\Psi K_S$ channel~\cite{paver}.

We think however that it is more convenient and more transparent to decouple the
different and intricated problems related to the determination of the UT. Fortunately,
such an attitude is simple to handle with, thanks to the CKM mechanism which predicts strong relations between CP-violating
and CP-conserving quantities: indeed the SM says that $\alpha$, $|V_{td}|$ and $|V_{ub}^\ast|$
are functions
of $(\rho,\eta)$~\footnote{For simplicity, we neglect the uncertainty on $V_{cb}$, and take $V_{ud}=V_{tb}=1$.}~\cite{wolf}:
\beqa
\alpha&=&\arg\left ( -\frac{1-\rho-i\eta}{\rho+i\eta} \right )\,,\\
|V_{td}|&=&\lambda|V_{cb}|\times|1-\rho-i\eta|\,,\\
|V_{ub}^\ast|&=&\lambda|V_{cb}|\times|\rho+i\eta|\,.
\eeqa
The above relations, inserted in Eqs.~(\ref{masterEq1}),~(\ref{masterEq2}) and~(\ref{masterEq3}), permit us to reexpress the latter as
equations in the $(\rho,\eta)$ variables, depending on the theoretical parameters
$|M^{(t)}|/|M^{(u)}|$, $|M^{(t)}|$ and $|M^{(u)}|$ respectively.
Indeed we define the following combinations of observables
\beq
D_c\equiv \sqrt{1-\adir^2}\cos2\alpha\eff\,,\ \ \ \ \ 
D_s\equiv \sqrt{1-\adir^2}\sin2\alpha\eff\,,
\eeq
and we introduce  $|M^{(t)}|$ and $|M^{(u)}|$, normalized to
$\sqrt{\B_{\pi^+\pi^-}}/|\lambda V_{cb}|$ (recall their definition~(\ref{TP}), and that the amplitudes squared are in ``units of two-body branching ratio'')
\beq\label{R_P&R_T}
\frac{R_P}{R_T}=\left|\frac{M^{(t)}}{M^{(u)}}\right|^2\,,\ \ \ \ \ 
R_P\equiv |\lambda V_{cb}|^2\frac{|M^{(t)}|^2}{\B_{\pi^+\pi^-}}\,,\ \ \ \ \ 
R_T\equiv |\lambda V_{cb}|^2\frac{|M^{(u)}|^2}{\B_{\pi^+\pi^-}}\,.
\eeq
Then Eqs.~(\ref{masterEq1}) and~(\ref{masterEq2}) are respectively equivalent to the
following degree-four polynomial equations, the first depending on $R_P/R_T$ only and the
second on $R_P$ only:
\beqa
&&(1-D_c)(1-\frac{R_P}{R_T})\,\rho^4+2(1-D_c)(1-\frac{R_P}{R_T})\,\rho^2\eta^2+(1-D_c)(1-\frac{R_P}{R_T})\,\eta^4 \nn\\
&&-2(1-D_c)(1-2\frac{R_P}{R_T})\,\rho^3-2D_s\,\rho^2\eta-2(1-D_c)(1-2\frac{R_P}{R_T})\,\rho\eta^2-2D_s\,\eta^3 \nn\\
&&+(1-D_c)(1-6\frac{R_P}{R_T})\,\rho^2+2D_s\,\rho\eta+[1+D_c-2(1-D_c)\frac{R_P}{R_T}]\,\eta^2\nn\\
&&+(1-D_c)\frac{R_P}{R_T}\,(4\rho-1)=0 \label{masterEq5}\,,
\eeqa
\beqa
&&(1-D_c)\,\rho^4+2(1-D_c-R_P)\,\rho^2\eta^2+(1-D_c-2R_P)\,\eta^4 \nn\\
&&-2(1-D_c)\,\rho^3-2D_s\,\rho^2\eta-2(1-D_c-2R_P)\,\rho\eta^2-2D_s\,\eta^3 \nn\\
&&+(1-D_c)\,\rho^2+2D_s\,\rho\eta+(1+D_c-2R_P)\,\eta^2=0 \label{masterEq6}\,.
\eeqa

When $M^{(t)}=0$ (the no-penguin case: $R_P=0$ and thus $\adir=0$), Eqs.~(\ref{masterEq5}) and~(\ref{masterEq6}) reduce to
\beq\label{cercleCarre}
\left[1-\cos2\alpha\eff\right]\left[\left(\rho-\frac{1}{2}\right)^2+\left(\eta-\frac{\sin2\alpha\eff}{2(1-\cos2\alpha\eff)}\right)^2-\frac{1}{2(1-\cos2\alpha\eff)}\right]^2\ =\ 0
\eeq
which is the equation squared of a circle. This circle is just, as expected, the one defined by $2\alpha=\arg[-(1-\rho-i\eta)/(\rho+i\eta)]=2\alpha\eff$ and can be obtained geometrically, by
using the definition of the UT and solving the equation $2\alpha={\rm C^{st}}$.
Actually, the sign of $\cos2\alpha\eff$ is not known and we get in fact two circles.
When $M^{(t)}\neq 0$, each of these two circles splits into two curves---this splitting is reminiscent of the $2\alpha-2\alpha\eff\to-(2\alpha-2\alpha\eff)$ ambiguity
of Eqs.~(\ref{masterEq1}) and~(\ref{masterEq2}): the no-penguin
case---the circle---appears as a double root of the general case---a degree-four
polynomial equation, as already noticed above when discussing Eq.~(\ref{masterEq1}).

Likewise Eq.~(\ref{masterEq3}) is equivalent to the following linear equation, depending on $R_T$ only:
\beq
\sqrt{1-D_c}\,(\rho-1) \pm \sqrt{2R_T-1+D_c}\,\,\eta=0 \label{masterEq7}\,.
\eeq
The $\pm$ sign is reminiscent of the $2\alpha\to-2\alpha$ ambiguity of Eq.~(\ref{masterEq3}).

As the parameter $R_T$ does not know much about the size of the penguin, the
no-penguin limit of Eq.~(\ref{masterEq7}) is not particularly interesting.

The important feature of Eqs.~(\ref{masterEq5}-\ref{masterEq7}) is that the parameters $R_P/R_T$, $R_P$ and $R_T$---defined by Eqs.~(\ref{TP}) and (\ref{R_P&R_T})---are {\it pure QCD quantities} times $|\lambda V_{cb}|^2/\B_{\pi^+\pi^-}$, i.e. they can be expressed as 
matrix elements of the Weak Effective Hamiltonian times known factors.

Thus the reader may choose a pure hadronic model to estimate $R_P/R_T$, $R_P$ or $R_T$, and report it in Eqs.~(\ref{masterEq5}),~(\ref{masterEq6}) or~(\ref{masterEq7}) respectively, then getting a polynomial equation which roots,
represented as curves, summarize the domain
in the $(\rho,\eta)$ plane which is allowed by the measurement of the time-dependent $B\to\pi\pi$
CP-asymmetry.
Some examples of this strategy are given in Section~\ref{K0pi+Section}, where we use
some phenomenological assumptions to estimate $R_P$, and in Section~\ref{previous},
where we suggest to improve the proposals of Fleischer and Mannel~\cite{FMzoology}
and Marrocchesi and Paver~\cite{paver} by solving the problem directly in the $(\rho,\eta)$ plane.
\section{Using Isospin Related Decays}
\label{pi0pi0Section}
In this section, we will assume:
\begin{itemize}
\item {\it SU(2) isospin symmetry of the strong interactions (Assumption~{\bf 2}).} 
It is well known that this flavour symmetry is indeed very good; in any case,
the violation of SU(2) should be completely negligible compared to the theoretical
errors discussed in this paper.
\end{itemize}
As the Effective Weak Hamiltonian is a linear combination of $\Delta I=1/2$ and $\Delta I=3/2$ operators, one has the triangular
relations (remember the notation~(\ref{TP}-\ref{TPpi+pi0}) and~(\ref{TP2}), and see~\cite{GL,LNQS}):
\beq\label{triangle}
T_{\pi^+\pi^0}=T+T_{\pi^0\pi^0}\,,\ \ \ \ \ P_{\pi^+\pi^0}=P+P_{\pi^0\pi^0}\,.
\eeq
As the QCD-penguins are pure $\Delta I=1/2$ amplitudes, the $P_{\pi^+\pi^0}$ amplitude come only from electroweak penguins. Thus we define $P_{\rm EW}=P_{\pi^+\pi^0}$ to get
\beqa
A(B^0\to\pi^+\pi^-)&=&e^{i\gamma}\,T+e^{-i\beta}P\,,\label{pi+pi-Isospin}\\
A(B^0\to\pi^0\pi^0)&=&\frac{1}{\sqrt{2}}\left[e^{i\gamma}\,T_{\pi^0\pi^0}+e^{-i\beta}(P_{\rm EW} -P)\right]\,,\label{pi0pi0Isospin}\\
A(B^+\to\pi^+\pi^0)&=&\frac{1}{\sqrt{2}}\left[e^{i\gamma}\,(T+T_{\pi^0\pi^0})+e^{-i\beta}P_{\rm EW}\right]\,.\label{pi+pi0Isospin}
\eeqa

The second assumption we will make here is:
\begin{itemize}
\item {\it Neglect of the electroweak penguin contributions in $B\to\pi^+\pi^-,\pi^\pm\pi^0,\pi^0\pi^0$ (Assumption~{\bf 5}).} That is $P_{\rm EW}=0$ in
Eqs.~(\ref{pi+pi-Isospin}-\ref{pi+pi0Isospin}). The problem with this approximation
arrives when
considering $B\to\pi^0\pi^0$, where, on naive grounds (short distance coefficients
and factorization of the matrix elements), the electroweak penguin---which is here
colour-allowed---is not particularly negligible. However the repercussion on the extraction of $\alpha$ is expected to
be negligible~\cite{Pew1,Pew2}, and in any case smaller than the gluonic penguin
effects. See also the discussion in \S~\ref{PewMasterEq}.
\end{itemize}

In the framework of these two assumptions, Gronau and London have shown that the
knowledge of the $B(\overline B)\to\pi^+\pi^-,\pi^0\pi^0,\pi^\pm\pi^0$ branching
ratios in addition to the time-dependent $B^0(t)\to\pi^+\pi^-$ CP-asymmetry leads
to the clean extraction of $\alpha$, up to discrete ambiguities. In \S~\ref{GronauLondon}, we reexpress the Gronau-London isospin analysis in our
language. In particular, we clarify the problem of the discrete ambiguities, which 
up to now has remained confused in the literature.

Unfortunately,
it is well known that the isospin study might be experimentally difficult to carry
out, if the
$B\to\pi^0\pi^0$ mode is as rare as expected because of colour-suppression. Therefore alternative methods have
to be developed.
\subsection{Two Upper Bounds on $|2\alpha-2\alpha\eff|$ from $B\to\pi^0\pi^0$}
In Ref.~\cite{GQbound} Grossman and Quinn have derived an interesting bound 
on the shift $|2\alpha-2\alpha\eff|$ (Eq. (2.12) of their paper) which takes in our notation the simple following form:
\beq\label{GQbound0}
|2\alpha-2\alpha\eff| \le \arccos\left ( 1-2\frac{\B_{\pi^0\pi^0}}{\B_{\pi^\pm\pi^0}} \right )\,.
\eeq
This bound derives from the isospin relations~(\ref{pi+pi-Isospin}-\ref{pi+pi0Isospin}) and from
the geometry of the Gronau-London triangle (see~\cite{GL,LNQS}) when the electroweak
penguin amplitude, i.e. $P_{\rm EW}$, is neglected. The physical 
meaning of this bound is simple: the $B\to\pi^0\pi^0$ branching ratio cannot
vanish exactly unless both the tree and the penguin amplitudes in $\pi^0\pi^0$ vanish, in which case $2\alpha=2\alpha\eff$ in $\pi^+\pi^-$.

As explained in Ref.~\cite{GQbound}, this bound is useful when the $B\to\pi^0\pi^0$
rate is too low, in which case only the average branching ratio $\B_{\pi^0\pi^0}$ (that can be obtained from {\it untagged} events only),
or even only an upper bound on this quantity, is available. Thus, either the 
$B\to\pi^0\pi^0$ channel is strong enough to allow a full isospin analysis, or the
rate is indeed very small and bounds the penguin-induced error on $\alpha$.

It is not difficult to derive
the bound~(\ref{GQbound0}) in an analytical way, different from the geometrical
approach of Ref.~\cite{GQbound}; here we give only the main line of the demonstration.
Neglecting the electroweak penguin ($P_{\rm EW}=0$) in Eqs.~(\ref{pi+pi-Isospin}-\ref{pi+pi0Isospin}) we can form the ratio $\B_{\pi^0\pi^0}/\B_{\pi^\pm\pi^0}$ and consider it as a function of the complex
parameter $T_{\pi^+\pi^0}$. Minimizing this ratio with respect to the latter parameter gives the inequality:
\beq
|P|^2\sin^2\alpha \,\le\, \frac{\B_{\pi^0\pi^0}}{\B_{\pi^\pm\pi^0}}\times\B_{\pi^+\pi^-}\,.
\eeq
Then $2\alpha$ is constrained thanks to Eq.~(\ref{masterEq2}):
\beq\label{GQbound}
|2\alpha-2\alpha\eff| \le \arccos \left [ \frac{1}{\sqrt{1-\adir^2}}
\left ( 1-2\frac{\B_{\pi^0\pi^0}}{\B_{\pi^\pm\pi^0}} \right )\right ]
\ \ \ \ \ \mbox{[assuming~{\bf  2} and~{\bf  5}].}
\eeq
As $0\le\adir\le 1$, the above bound is slightly more stringent than the bound~(\ref{GQbound0}), and reduce to the latter when $\adir=0$.

Under the same isospin symmetry and neglect of electroweak penguin hypotheses, it is straightforward to derive another similar bound, not given in the original paper~\cite{GQbound}. Indeed, using the general bounds~(\ref{galBounds}) for the penguin in Eq.~(\ref{TPpi0pi0}) gives simply:
\beq
|P|^2\sin^2\alpha \,\le\, 2\B_{\pi^0\pi^0}\,,
\eeq
where the factor two is related to a Clebsh-Gordan coefficient, i.e. to the
wave function of the $\pi^0$ meson and the Bose symmetry. Once again we use
Eq.~(\ref{masterEq2}) to get:
\beq\label{pi0pi0Bound}
|2\alpha-2\alpha\eff| \le \arccos \left [ \frac{1}{\sqrt{1-\adir^2}}
\left ( 1-4\frac{\B_{\pi^0\pi^0}}{\B_{\pi^+\pi^-}} \right )\right ]
\ \ \ \ \ \mbox{[assuming~{\bf  2} and~{\bf  5}].}
\eeq

To which extent the Grossman-Quinn bound~(\ref{GQbound}) is better than the
bound~(\ref{pi0pi0Bound}) or {\it vice-versa} depends on the actual values of the branching
ratios $2\B_{\pi^\pm\pi^0}$ vs. $\B_{\pi^+\pi^-}$: in fact, neglecting penguin and colour-suppressed
contributions would lead to the equality $2\B_{\pi^\pm\pi^0}=\B_{\pi^+\pi^-}$, while the factorization assumption, predicting a
constructive interference between the colour-allowed and colour-suppressed contributions in
the $B^\pm\to\pi^\pm\pi^0$ channel, tends to favour~(\ref{GQbound}) compared to~(\ref{pi0pi0Bound})~\footnote{Grossman and Quinn give another bound depending on both $\B_{\pi^0\pi^0}/\B_{\pi^\pm\pi^0}$ and
$\B_{\pi^0\pi^0}/\B_{\pi^+\pi^-}$ (eq. (2.15) of their paper~\cite{GQbound}). As it is more complicated and presumably numerically similar, we do not report it here.}. 
A technical advantage of the bound~(\ref{pi0pi0Bound}) over~(\ref{GQbound}) is that
it does not require the measurement of $\B_{\pi^\pm\pi^0}$, which may be less well
measured than $\B_{\pi^+\pi^-}$ because of the necessary $\pi^0$ detection, and because it is expected that $\B_{\pi^\pm\pi^0}<\B_{\pi^+\pi^-}$.
Note in passing that the Grossman-Quinn bound
follows from the isospin constraints on both tree and penguin amplitudes, while our
bound comes only from the isospin constraints on the penguin. Of course it can be
checked that the two bounds are fully compatible, in the sense that when the
bound~(\ref{GQbound}) is saturated then the bound~(\ref{pi0pi0Bound}) is automatically
satisfied and {\it vice-versa}.

As shown by Grossman and Quinn~\cite{GQbound}, and as redemonstrated
for consistency in Appendix~\ref{2aeffbar}, the above bounds imply weaker ones, obtained by replacing $\adir$ by zero and $2\alpha\eff$ by $2\overline{\alpha}\eff$ where $2\overline{\alpha}\eff$ is defined by Eq.~(\ref{2aeffbarDef}). Thus we have:
\beqa
&&|2\alpha-2\overline{\alpha}\eff| \le \arccos  \left ( 1-2\frac{\B_{\pi^0\pi^0}}{\B_{\pi^\pm\pi^0}} \right )\,, \\
&&|2\alpha-2\overline{\alpha}\eff| \le \arccos  \left ( 1-4\frac{\B_{\pi^0\pi^0}}{\B_{\pi^+\pi^-}} \right )
\ \ \ \ \ \mbox{[assuming~{\bf  2} and~{\bf  5}].}
\eeqa
As already stressed, the advantage in using $2\overline{\alpha}\eff$ instead of $2\alpha\eff$  is that
the former angle follows directly (up to a twofold ambiguity) from the $\sin\Delta m\,t$ term in the time-dependent CP-asymmetry~(\ref{timeCP2}) independently of 
$\adir$, and thus does not require the measurement of the latter.
\section{Using SU(3) Related Decays}
\label{SU3section}
In this section we will assume a larger flavour symmetry, namely
\begin{itemize}
\item
{\it SU(3) flavour symmetry of the strong interactions (Assumption~{\bf 3}).} One could argue that such an assumption should not be too bad in energetic two-body decays, although we know that a typical SU(3) breaking quantity is $|f_K-f_{\pi}|/f_{\pi}\sim 23\%$. Actually,
our present knowledge does not permit a reliable quantitative estimate of such
a symmetry breaking in $B$ decays, especially for the penguin amplitudes that we are interested in.
In any case, our understanding of this problem is expected to improve with both
theoretical and experimental progresses.
\end{itemize}
\subsection{An Upper Bound on $|2\alpha-2\alpha\eff|$ from $B\to K^0\overline{K^0}$}
Very similarly to the isospin analysis and the $B\to\pi^0\pi^0$ case, it is possible to derive a bound
on $|2\alpha-2\alpha\eff|$ depending on $\BR(B\to K^0\overline{K^0})$. Indeed Buras
and Fleischer~\cite{burasK0K0b} have proposed a SU(3) analysis which rely on the
measurement of the time-dependent CP-asymmetry in the $B\to K^0\overline{K^0}$ channel to disentangle the penguin effects in $B\to\pi^+\pi^-$ (see \S~\ref{BF}). However the $B\to K^0\overline{K^0}$ channel is a pure $b\to d$ penguin process, and its rate is presumably rather small ($10^{-7}$--$10^{-6}$). Nevertheless, due to the bounds~(\ref{galBounds}), $\B_{K^0\overline{K^0}}$ cannot vanish unless
both $T_{K^0\overline{K^0}}$ and $P_{K^0\overline{K^0}}$ vanish in Eq.~(\ref{TPK0K0b}). Thus either $\B_{K^0\overline{K^0}}$ is large enough to do the Buras-Fleischer analysis, or it is vanishingly small and one expects that  $|2\alpha-2\alpha\eff|$ is constrained by $\B_{K^0\overline{K^0}}$ thanks to the
SU(3) symmetry. Similarly to the case of $\pi^0\pi^0$, an upper bound on $\B_{K^0\overline{K^0}}$ is sufficient to get constraints on $\alpha$.

In addition to the SU(3)
flavour symmetry introduced above, we need the following assumption:
\begin{itemize}
\item {\it Neglect of the electroweak penguin contributions in $B\to\pi^+\pi^-,K^0\overline{K^0}$ (Assumption~{\bf 5}).} In Ref.~\cite{burasK0K0b}, Buras and Fleischer argue that this approximation is better than its
equivalent in the Gronau-London construction. Indeed, in the latter case the 
electroweak penguin is colour-allowed, while in the
present case it is colour-suppressed. However one has to remind
that FSI effects may invalidate the notion of colour-suppression~\cite{neubert}.  
\end{itemize}
Within the above assumptions, we have
\beq\label{su3K0K0b}
|P|=|P_{K^0\overline{K^0}}|\,.
\eeq
Then we repeat the demonstration
given above for the $B\to\pi^0\pi^0$ channel to obtain from Eqs.~(\ref{TPK0K0b}),~(\ref{galBounds}), and~(\ref{masterEq2}):
\beq\label{K0K0bBound}
|2\alpha-2\alpha\eff| \le \arccos \left [ \frac{1}{\sqrt{1-\adir^2}}
\left ( 1-2\frac{\B_{K^0\overline{K^0}}}{\B_{\pi^+\pi^-}} \right )\right ]
\ \ \ \ \ \mbox{[assuming~{\bf  3} and~{\bf  5}].}
\eeq
Likewise (Appendix~\ref{2aeffbar}), under the same hypotheses there is a bound independent of $\adir$, using the angle $2\overline{\alpha}\eff$:
\beq\label{K0K0bBound2}
|2\alpha-2\overline{\alpha}\eff| \le \arccos  \left ( 1-2\frac{\B_{K^0\overline{K^0}}}{\B_{\pi^+\pi^-}} \right )
\ \ \ \ \ \mbox{[assuming~{\bf  3} and~{\bf  5}].}
\eeq

Hence, analogously to the isospin analysis and the bounds~(\ref{GQbound}) and~(\ref{pi0pi0Bound}), our
bounds~(\ref{K0K0bBound}-\ref{K0K0bBound2}) may be useful when the $B\to K^0\overline{K^0}$ channel is too rare to achieve the full Buras-Fleischer analysis, and thus only the value of $\B_{K^0\overline{K^0}}$, or even an upper limit on
this branching ratio, is available.
\subsection{An Upper Bound on $|2\alpha-2\alpha\eff|$ from $B\to K^\pm\pi^\mp$}
\label{K+pi-Section}
It has been known for a long time that the $B\to K\pi$ decays can help the extraction of
$\alpha$ from the time-dependent $B\to\pi\pi$ CP-asymmetry by constraining
the penguin amplitudes~\cite{silvWolf}. Indeed,
the latter are doubly-Cabibbo-enhanced by the ratio $|V_{cs}V_{cb}^\ast/(V_{us}V_{ub}^\ast)|$ with respect to the tree in these $K\pi$ decays. However,
in addition to the unavoidable SU(3) assumption, people are often lead to neglect annihilation diagrams and/or
electroweak penguins and/or $u,c$-penguins and/or Final State Interaction (FSI) in
expressing the $B\to\pi\pi$ amplitudes in terms of the $B\to K\pi$ ones~\cite{silvWolf,gronauSU3}. Such ill-defined approximations
have been questioned in the recent literature~\cite{neubert,polemiqueKpi} in connection with
the so-called Fleischer-Mannel bound on $\sin^2\gamma$~\cite{FMbound}. Here however, in addition to SU(3), we will only use the following approximation in comparing Eqs.~(\ref{TP})
and~(\ref{TPK+pi-}):
\begin{itemize}
\item
{\it Neglect of the OZI-suppressed annihilation penguin diagrams (Assumption~{\bf  4}).} The topology of these
diagrams is represented in Fig.~\ref{fig:penguins}. We will need to neglect these diagrams
only for the $P$-amplitude, i.e. when the quark in the loop is a $t$ or a $c$ (recall Eq.~(\ref{TP})). When the flavour in the loop is $t$, the suppression is perturbative,
due to a linear combination of short-distance Wilson coefficients which is $\sim\alpha_s^2(m_b)$; on the contrary, the same diagram with a $c$ quark is
non-perturbatively suppressed by the OZI-rule~\cite{rome1}. In addition these
diagrams are usually expected
to be suppressed by the annihilation topology. Thus they are probably very small and negligible compared to the SU(3)-induced theoretical error. 
\begin{figure}[ht]
\centerline{\epsfxsize=0.49\textwidth\epsfbox{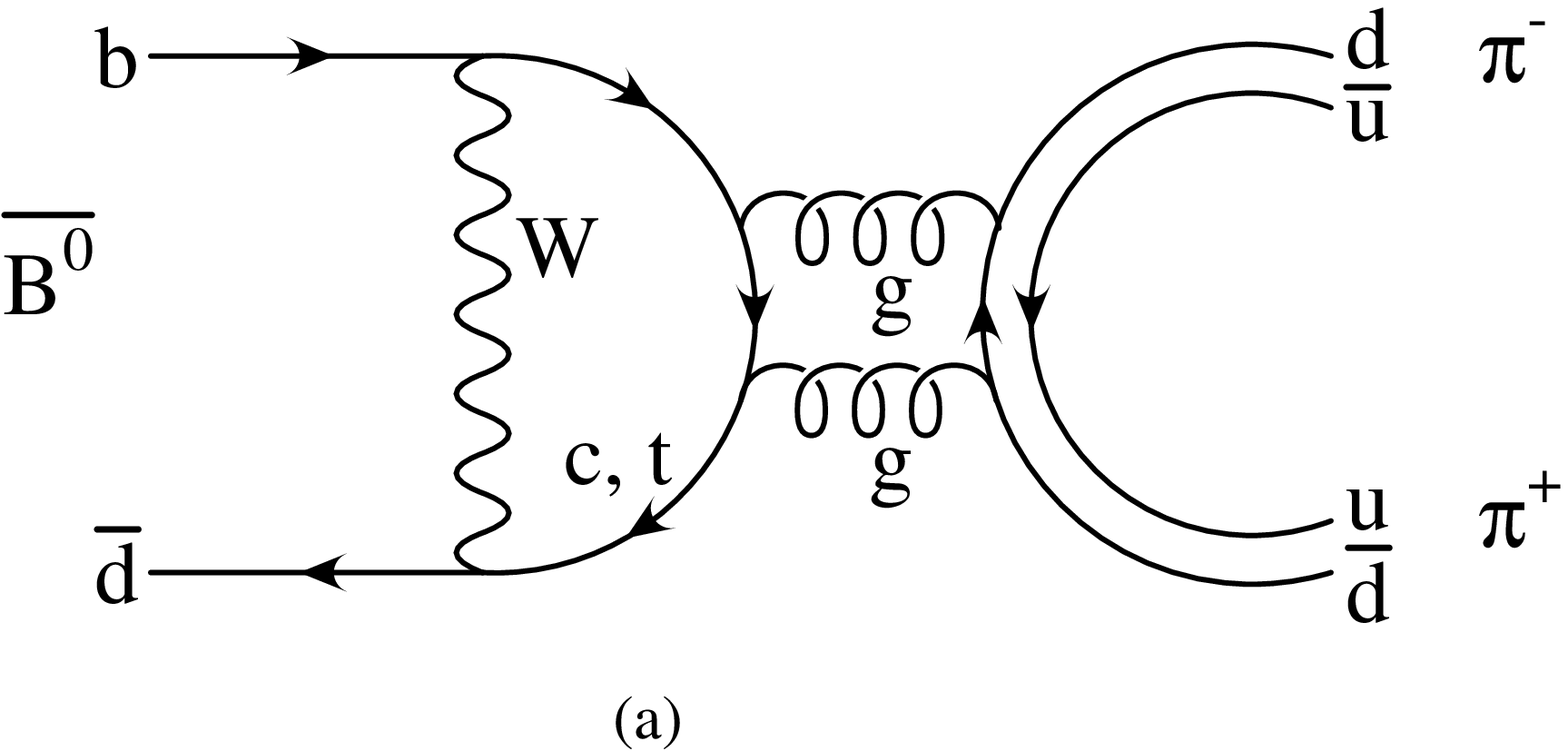}\hspace{0.01\textwidth}
\epsfxsize=0.49\textwidth\epsfbox{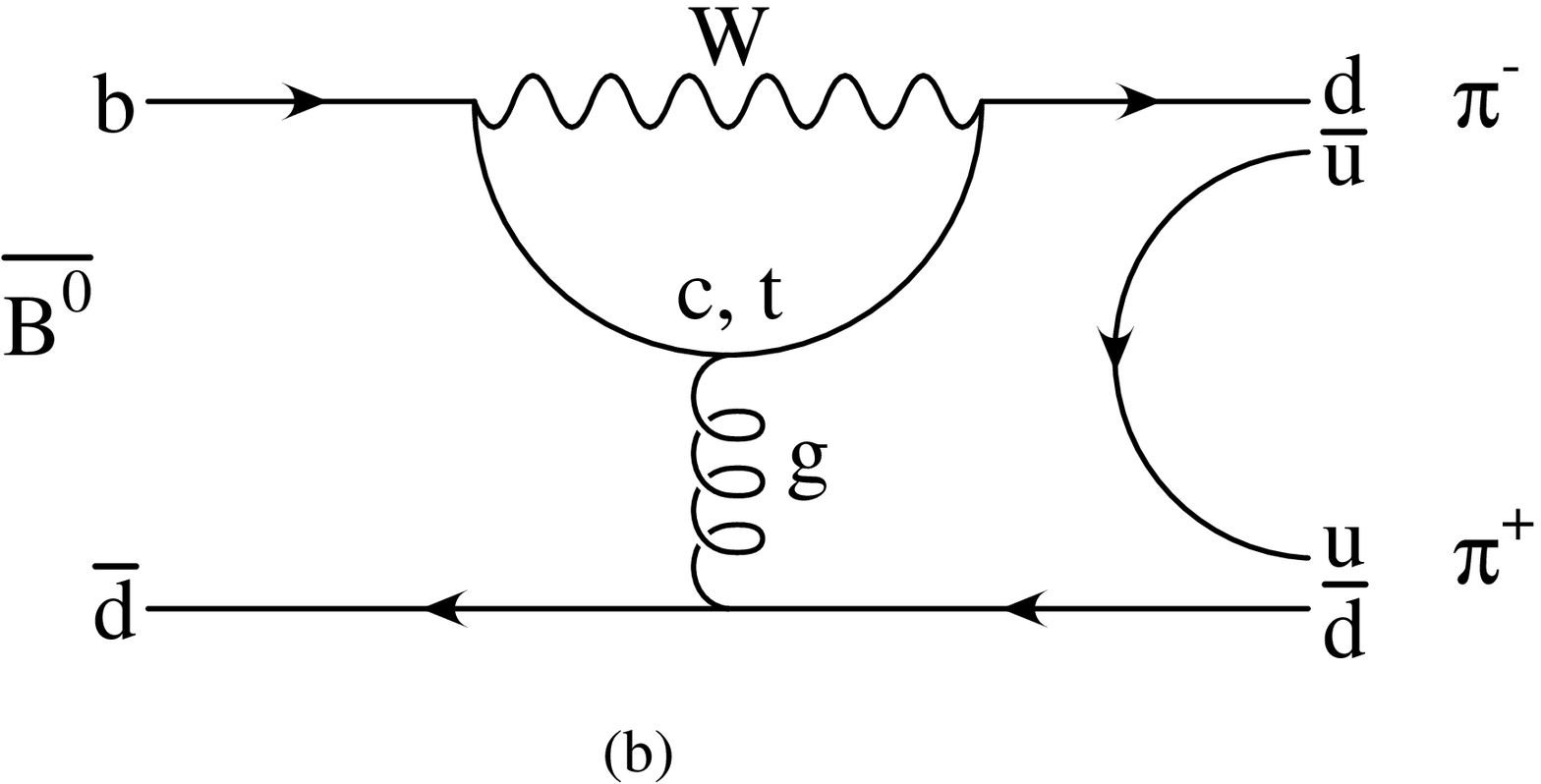}}
\caption{\it {\rm(a)} OZI-suppressed annihilation penguin diagram. This diagram is OZI-suppressed and is neglected within Assumption~{\bf  4} because it does not contribute to $B\to K^\pm\pi^\mp$.
{\rm(b)} Non-OZI-suppressed annihilation penguin diagram. This diagram is not OZI-suppressed, although it has annihilation topology. \underbar{It is not neglected} within Assumption~{\bf  4} because it contributes to both $B\to\pi^+\pi^-$ and $B\to K^\pm\pi^\mp$.}\label{fig:penguins}
\end{figure}
\end{itemize}

In particular, {\it we do not neglect the electroweak 
penguin} as it produces the same amplitude in $B\to K^\pm\pi^\mp$ and $B\to\pi^+\pi^-$,
assuming SU(3) and neglecting OZI-suppressed penguins.
Then we get simply from Eqs.~(\ref{TP}),~(\ref{TP2}) and~(\ref{TPK+pi-})
\beq\label{su3K+pi-}
|P|=\left|\frac{V_{td}V_{tb}^\ast}{V_{cs}V_{cb}^\ast}\right|\times|P_{K^+\pi^-}|
=\lambda\frac{\sin\gamma}{\sin\alpha}\,|P_{K^+\pi^-}|
\eeq
where the geometry of the UT has been used in writing $|V_{td}/V_{cb}^\ast|$.
From Eqs.~(\ref{TPK+pi-}) and~(\ref{galBounds}), we get the following bound on $|P_{K^+\pi^-}|$:
\beq\label{FMbound}
|P_{K^+\pi^-}|^2\sin^2\gamma \,\le\, \B_{K^\pm\pi^\mp}\,.
\eeq
Combining it with Eqs.~(\ref{masterEq2}) and~(\ref{su3K+pi-}) we obtain~\footnote{This is a somewhat miraculous feature of the SM: $\sin\gamma$ cancels
between Eqs.~(\ref{su3K+pi-}) and~(\ref{FMbound}).}
\beq\label{K+pi-Bound}
|2\alpha-2\alpha\eff| \le \arccos \left [ \frac{1}{\sqrt{1-\adir^2}}
\left ( 1-2\lambda^2\,\frac{\B_{K^\pm\pi^\mp}}{\B_{\pi^+\pi^-}} \right )\right ]
\ \ \ \ \ \mbox{[assuming~{\bf  3} and~{\bf  4}].}
\eeq
Thus this bound is a quantitative realization of the well-known fact that the penguin
in $\pi\pi$ is $\lambda$-suppressed with respect to the penguin in $K\pi$: if
$\lambda^2\B_{K^\pm\pi^\mp}/\B_{\pi^+\pi^-}$ is not too large, this means that the penguin
cannot be too large in $B\to\pi\pi$. Note that similarly to the $B\to\pi^0\pi^0$
and $B\to K^0\overline{K^0}$ channels, {\it we have not assumed penguin dominance
in $B\to K\pi$} although of course the bound should be more interesting when the penguin
amplitude really dominates in the latter decay.

From the experimental point of view, the bound that we have found in Eq.~(\ref{K+pi-Bound}) should be considerably less affected by statistical uncertainties than the bounds~(\ref{GQbound}), (\ref{pi0pi0Bound}) and~(\ref{K0K0bBound}). Indeed, rather than measuring the branching ratio of
very rare decays such as $B\to\pi^0\pi^0$ or $B\to K^0\overline{K^0}$, the use of the bound~(\ref{K+pi-Bound})
needs to know $\B_{K^\pm\pi^\mp}$ which is ${\cal O}(10^{-5})$. For this reason the CLEO data~(\ref{CLEO}) already help to give a very interesting and non trivial estimation of the r.h.s. of~(\ref{K+pi-Bound}).
Indeed~(\ref{K+pi-Bound}) imply $|2\alpha-2\alpha\eff| \le \arccos( 1-2\lambda^2\B_{K^\pm\pi^\mp}/\B_{\pi^+\pi^-})$ and from~(\ref{CLEO}) we have $0.81\times 10^{-5}<\B_{K^\pm\pi^\mp}< 2.0\times 10^{-5}$ and $\B_{\pi^+\pi^-}< 0.84\times 10^{-5}$ at 90\% C.L.. Assuming furthermore
$0.4\times 10^{-5}<\B_{\pi^+\pi^-}$---otherwise the study of CP-violation in the $\pi\pi$ channel would
be very difficult, independently of penguins---we obtain the bound
\beq\label{K+pi-BoundNum}
|2\alpha-2\alpha\eff|\le\Delta\,,\ \ \ \mbox{with}\ \ \ 25^\circ<\Delta<59^\circ\,.
\eeq
Thus, although these data indicate that the extraction of $\alpha$ will not be an easy task,
they are still compatible with a relatively small penguin-induced theoretical error. We would like to stress also that to our knowledge, this is the first time that a numerical upper bound on the theoretical error on $\alpha$ 
is given rather model-independently, with only mild theoretical assumptions and 
before the experimental value of $\alpha\eff$ is available by itself. It is expected that
experiment will give an accurate determination of the r.h.s. of~(\ref{K+pi-Bound})
quite soon. Unless we are unlucky and $\B_{\pi^+\pi^-}$ is much smaller than
expected, the theoretical errol on $\alpha$ constrained by the bound~(\ref{K+pi-Bound}) should not excess $\sim 30^\circ$ while it can be as small
as $\sim 10^\circ$. In comparison, the
current knowledge of $\alpha$ is roughly $40^\circ<\alpha<140^\circ$~\footnote{We stress that although there are presently very weak constraints on $\sin2\alpha$~\cite{LAL},
this is not the case for $\alpha$ itself.}.

Finally one has again a bound independent of $\adir$ where $2\overline{\alpha}\eff$ is involved:
\beq
|2\alpha-2\overline{\alpha}\eff| \le \arccos  \left ( 1-2\lambda^2\,\frac{\B_{K^\pm\pi^\mp}}{\B_{\pi^+\pi^-}} \right )
\ \ \ \ \ \mbox{[assuming~{\bf  3} and~{\bf  4}].}
\eeq

Let us note in passing that the inequality~(\ref{FMbound}), together with the assumption $|P_{K^+\pi^-}|^2=\BR(B^\pm\to K\pi^\pm)\equiv \B_{K\pi^\pm}$ (that we will use in 
Section~\ref{K0pi+Section}), leads to the Fleischer-Mannel bound on $\sin^2\gamma$~\cite{FMbound}.
However, the bound~(\ref{FMbound}) is exact independently of the assumption $|P_{K^+\pi^-}|^2=\B_{K\pi^\pm}$.
\subsection{Upper Bounds on $|2\alpha-2\alpha\eff|$: Numerical Examples}
\label{numericalBound}
In Appendix~\ref{model}, we define a typical set of parameters for the quantities
involved in the channels that we are interested in. This set of parameters, compatible with the CLEO data~(\ref{CLEO}), allows
to compute the various observables (branching ratios and CP-asymmetries), and in particular to estimate numerically the
bounds we have derived until now:
\beq\label{b1}
|2\alpha-2\alpha\eff| \le \arccos \left [ \frac{1}{\sqrt{1-\adir^2}}
\left ( 1-2\frac{\B_{\pi^0\pi^0}}{\B_{\pi^\pm\pi^0}} \right )\right ]
=33.4^\circ
\eeq
\beq\label{b2}
|2\alpha-2\alpha\eff| \le \arccos \left [ \frac{1}{\sqrt{1-\adir^2}}
\left ( 1-4\frac{\B_{\pi^0\pi^0}}{\B_{\pi^+\pi^-}} \right )\right ]
=40.3^\circ
\eeq
\beq\label{b3}
|2\alpha-2\alpha\eff| \le \arccos \left [ \frac{1}{\sqrt{1-\adir^2}}
\left ( 1-2\frac{\B_{K^0\overline{K^0}}}{\B_{\pi^+\pi^-}} \right )\right ]
=27.1^\circ
\eeq
\beq\label{b4}
|2\alpha-2\alpha\eff| \le \arccos \left [ \frac{1}{\sqrt{1-\adir^2}}
\left ( 1-2\lambda^2\,\frac{\B_{K^\pm\pi^\mp}}{\B_{\pi^+\pi^-}} \right )\right ]
=29.9^\circ
\eeq
The true value being $2\alpha-2\alpha\eff=+26.7^\circ$ (Appendix~\ref{model}),
the bound~(\ref{b3}) is very close to be saturated for this set of parameters.
Note also that the bounds~(\ref{b1}-\ref{b4}) are numerically close one from one another, that just follows from our set of parameters and needs not to be
true in general. As said above, it may happen in practice that the experiment gives only an upper
bound on the suppressed channels $\B_{\pi^0\pi^0}$ and $\B_{K^0\overline{K^0}}$,
in which case the bound~(\ref{b4}) will certainly be more informative as $\B_{K^\pm\pi^\mp}$ is already measured and hopefully the ratio $\B_{K^\pm\pi^\mp}/\B_{\pi^+\pi^-}$ will be known with high accuracy very
soon.

\begin{figure}[ht]
\centerline{\epsfxsize=0.57\textwidth\epsfbox{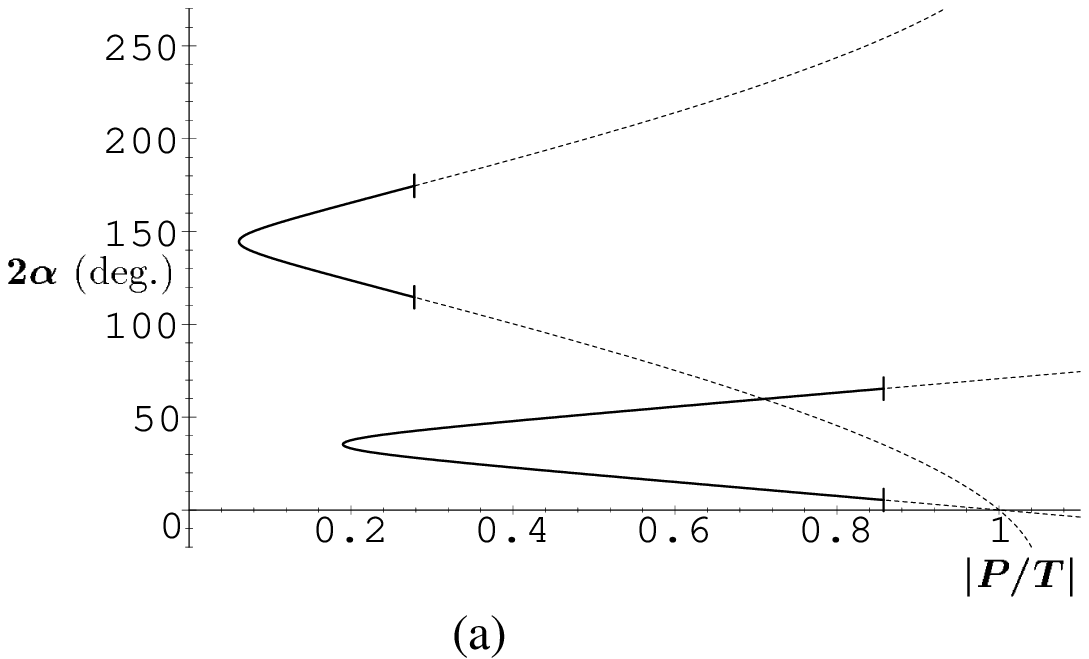}\hspace{0.01\textwidth}
\epsfxsize=0.4\textwidth\epsfbox{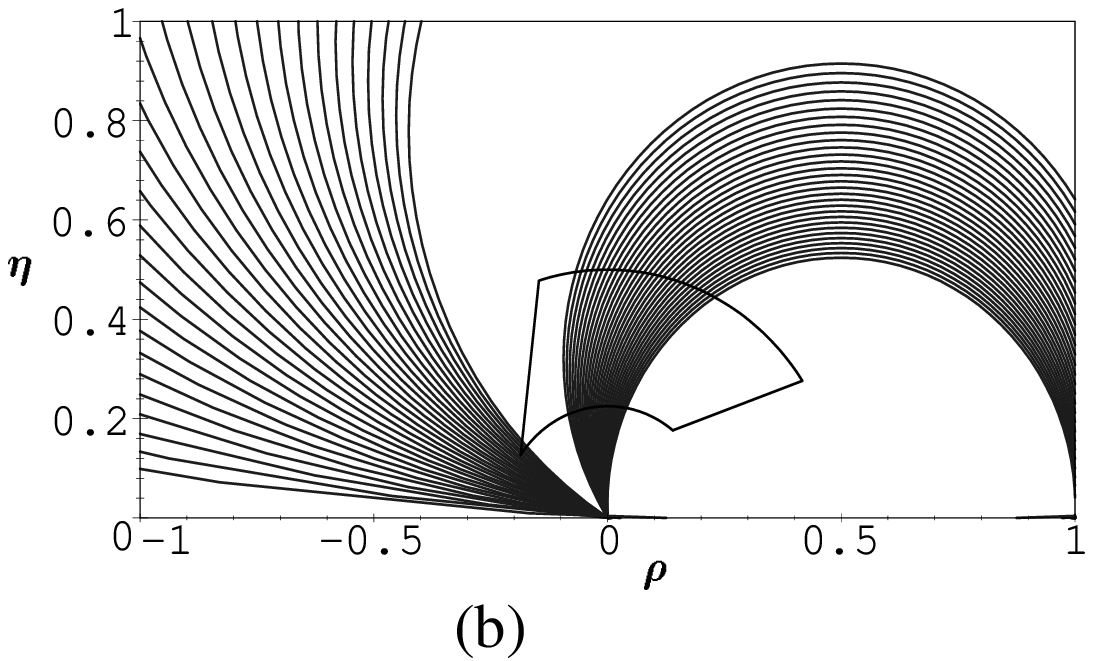}}
\caption{\it {\rm(a)} The bound $|2\alpha-2\alpha\eff|\le \Delta$ in the $(|P/T|,2\alpha)$
plane, obtained as explained
in the text, for the numerical example $\adir=0.12$, $\sin2\alpha\eff=0.58$ (see Appendix~\ref{model}) and $\Delta=30^\circ$ (solid curves limited by the small dashes). {\rm(b)} The same bound 
in the $(\rho,\eta)$ plane, obtained by plotting the circle~(\ref{cercleAlpha}) for
$2\alpha$ varying in the interval $[2\alpha\eff-\Delta,2\alpha\eff+\Delta]$. There
are two families of circles corresponding to the two possible signs for
$\cos2\alpha\eff$. In the background is shown a crude representation of the early-1998 allowed domain~\cite{LAL2}.}
\label{fig:borne}
\end{figure}

In any case and for the illustrative purpose, we will examine the case of the bound
$|2\alpha-2\alpha\eff|\le 30^\circ$, which is in the ball-park of~(\ref{b1}-\ref{b4}). On Fig.~\ref{fig:borne} we show the constraints of such
a bound in the $(|P/T|,2\alpha)$ plane, and in the $(\rho,\eta)$ plane. The latter
are obtained by plotting the circle defined by $2\alpha={\rm C^{st}}$, which equation is (cf. also Eq.~(\ref{cercleCarre})):
\beq\label{cercleAlpha}
\left(\rho-\frac{1}{2}\right)^2+\left(\eta-\frac{\sin2\alpha}{2(1-\cos2\alpha)}\right)^2=\frac{1}{2(1-\cos2\alpha)}\,.
\eeq
We let $2\alpha$ vary in the interval $[2\alpha\eff-30^\circ,2\alpha\eff+30^\circ]$ that is
consistent with the bound.

Thus the bounds~(\ref{b1}-\ref{b4}) should give important and rather safe information
on the angle $2\alpha$ as it is apparent on Fig.~\ref{fig:borne}.  Even if the penguin-induced
error on $\alpha$ may be large, {\it it is bounded by theoretical arguments,} which
is already an important statement in view of the possible tests of the consistency
of the SM.
\subsection{A Lower Bound on $|2\alpha-2\alpha\eff|$ from $B\to K^\pm\pi^\mp$}
\label{lowerBound}
It is clear from the examples in \S~\ref{numericalBound} that a {\it lower}
bound on $|2\alpha-2\alpha\eff|$ would be a valuable information: it would permit
to eliminate some region around $2\alpha=2\alpha\eff$ and to get four separate intervals
for $2\alpha$ (cf. Fig~\ref{fig:alpha}) instead of the two big ones represented on Fig.~\ref{fig:borne}. Thus one may
look for a lower bound on the absolute magnitude of the penguin. However, without
any further theoretical assumptions (see Section~\ref{K0pi+Section}), such a lower
bound cannot be obtained using branching ratios only (for example the bound discussed in \S~\ref{magnitude} is not theoretically justified). Conversely,
using direct CP-asymmetry in the $B^\pm\to K^\pm\pi^\mp$ decay, it
is possible to get a lower bound on $|P|$, as well as a slightly improved upper bound with respect to the bound~(\ref{K+pi-Bound}). The idea is the following: if a direct CP-\-asymmetry
in the $B\to K^\pm\pi^\mp$ channel is detected, then it proves that this mode is
fed by both tree and penguin contributions. As the latter is related by SU(3) to
the penguin in the $B\to\pi^+\pi^-$ channel, and thus to the penguin-induced shift
$|2\alpha-2\alpha\eff|$, one gets a lower bound on this quantity~\footnote{Note that
a non-vanishing direct CP-asymmetry in the $\pi^+\pi^-$ channel gives already a lower bound
on the penguin contributions through Eq.~(\ref{PsurTintervalle}). However the
saturation of the latter bound imply only $2\alpha=2\alpha\eff$. Thus one should look for a lower bound on the penguin that has to be stronger than Eq.~(\ref{PsurTintervalle}).}.

Analogously to the derivation of Eq.~(\ref{masterEq2}), we get from Eq.~(\ref{TPK+pi-})
\beq\label{Ps}
2\,|P_{K^+\pi^-}|^2\sin^2\gamma=\left [ 1-\sqrt{1-\left[\adir^{K\pi}\right]^2}\cos\zeta
\right ] \B_{K^\pm\pi^\mp}
\eeq
where
\beq
{\adir^{K\pi}}=\frac{\BR(B^0\to K^+\pi^-)-\BR(\overline{B^0}\to K^-\pi^+)}
{\BR(B^0\to K^+\pi^-)+\BR(\overline{B^0}\to K^-\pi^+)}
\eeq
should be relatively easy to measure for this self-tagging mode,
and $\zeta$ is a useful short-hand for the phase
\beq
\zeta=2\gamma-\arg\left [ A(B^0\to K^+\pi^-)A^\ast(\overline{B^0}\to K^-\pi^+)\right ]\,.
\eeq
Then, the inequality $|\cos\zeta|\le 1$ together with
Eqs.~(\ref{su3K+pi-}) and~(\ref{Ps}) imply
\beq\label{K+Pi-boundP}
\lambda^2\left(1-\sqrt{1-\left[\adir^{K\pi}\right]^2}\right)\,\B_{K^\pm\pi^\mp} \,\le\, 2\,|P|^2\sin^2\alpha \,\le\, \lambda^2\left(1+\sqrt{1-\left[\adir^{K\pi}\right]^2}\right)\,\B_{K^\pm\pi^\mp}
\eeq
and from Eq.~(\ref{masterEq2})
\beq\label{K+Pi-Ubound}
|2\alpha-2\alpha\eff| \le \arccos \left \{ \frac{1}{\sqrt{1-\adir^2}}
\left [ 1-\lambda^2\left ( 1+\sqrt{1-\left[\adir^{K\pi}\right]^2}\right ) \,\frac{\B_{K^\pm\pi^\mp}}{\B_{\pi^+\pi^-}} \right ]\right \}\,,
\eeq
\beqa\label{K+Pi-Lbound}
\arccos \left \{ \frac{1}{\sqrt{1-\adir^2}}
\left [ 1-\lambda^2\left ( 1-\sqrt{1-\left[\adir^{K\pi}\right]^2}\right ) \,\frac{\B_{K^\pm\pi^\mp}}{\B_{\pi^+\pi^-}} \right ]\right \}
\le |2\alpha-2\alpha\eff|\,,
\eeqa
{\begin{flushright}[assuming~{\bf  3} and~{\bf  4}].\end{flushright}}

Note that if
\beq\label{condition}
\lambda^2\left ( 1-\sqrt{1-\left[\adir^{K\pi}\right]^2}\right )\,\B_{K^\pm\pi^\mp} \,\le\, \left ( 1-\sqrt{1-\adir^2}\right )\,\B_{\pi^+\pi^-}
\eeq
the SU(3) lower bound in~(\ref{K+Pi-boundP}) is useless as it is automatically verified thanks to Eq.~(\ref{masterEq1}) and the (exact) bound~(\ref{PsurTintervalle}). Thus this lower bound
is only useful in the configuration where the direct CP-asymmetry is very small in the
$B\to\pi^+\pi^-$ channel ($\adir\to 0$) but large in the $B\to K^\pm\pi^\mp$ one (it becomes trivial in the limit $\adir^{K\pi}\to 0$), in which case the inequality~(\ref{condition}) is not verified. As an example, it can be checked that the set of parameters
defined in Appendix~\ref{model} verifies~(\ref{condition}). However, keeping the same 
branching ratios and choosing the parameters such as $\adir=0$ and $\adir^{K\pi}=0.5$,
the bound~(\ref{K+Pi-Lbound}) is not trivial:
\beq
8^\circ \le |2\alpha-2\alpha\eff|\,,
\eeq
while the bound~(\ref{K+Pi-Ubound}) represents only a tiny improvement over~(\ref{K+pi-Bound}).

Actually, one easily obtains similar lower bounds from the two previously studied
channels, namely $B\to\pi^0\pi^0$, $B\to K^0\overline{K^0}$.
However, the experimental detection of direct CP-violation in these suppressed
channels may be a difficult task. Should it be feasible, one may do the full
Gronau-London and/or Buras-Fleischer analyses (see Section~\ref{previous}).
\section{Using the $B^{\pm}\to K\pi^{\pm}$ Decay to Determine $R_P$ with Further Assumptions}
\label{K0pi+Section}
In this section, in addition to the hypothesis made in \S~\ref{K+pi-Section} (Assumptions~{\bf  3} and~{\bf  4}), we will assume more specifically that the two following approximations
hold (to an accuracy to be determined) in Eqs.~(\ref{TPK+pi-}-\ref{TPK0pi+}):
\begin{itemize}
\item
{\it Isospin symmetry and neglect of electroweak penguin contributions in $B\to K^\pm\pi^\mp,K\pi^\pm$ (Assumptions~{\bf  2} and~{\bf  5}).} Note that the isospin symmetry is a consequence of the already assumed larger
SU(3) symmetry. Neglecting the electroweak penguin, which is here colour-suppressed~\cite{Pew1}, we are allowed
to write $P_{K^+\pi^-}=-P_{K^0\pi^+}$~\cite{LNQS}.
\item
{\it Neglect of the $V_{us}V_{ub}^\ast$ contribution to the $B^+\to K^0\pi^+$ amplitude (Assumption~{\bf  6}).} That is, $T_{K^0\pi^+}=0$. Using a diagrammatic decomposition of the amplitude
, we have $T_{K^0\pi^+}=|V_{us}V_{ub}^\ast|(M_a+M_u-M_t)$ and $P_{K^0\pi^+}=|V_{cs}V_{cb}^\ast|(M_c-M_t)$, where $M_a$ is the tree
annihilation amplitude and $M_u$ (resp. $M_c$ resp. $M_t$) is the $u$- (resp.
$c$- resp. $t$-) penguin. Note that $T_{K^0\pi^+}$ is suppressed by $|V_{us}V_{ub}^\ast|/|V_{cs}V_{cb}^\ast|\sim 2\times 10^{-2}$ compared to the dominant amplitude $P_{K^0\pi^+}$. Thus we have presumably $|V_{us}V_{ub}^\ast||M_u-M_t| \ll |V_{cs}V_{cb}^\ast||M_c-M_t|$, and it
is often assumed that annihilation processes are negligible due to form-factor
suppression~\cite{gronauSU3}, which then lead to $|T_{K^0\pi^+}|\ll |P_{K^0\pi^+}|$.
\end{itemize}

It is clear that Assumption~{\bf  6} is on weaker grounds than the others made until now~\footnote{In particular, it implies a non-trivial relation
between FSI phases~\cite{wolfKpi}.}. Accepting it nevertheless, one is lead to many applications~\cite{gronauSU3,FMzoology} among which the most recent one is the
Fleischer-Mannel bound~\cite{FMbound}
\beq\label{FMbound2}
\sin^2\gamma \,\le\, \frac{\B_{K^\pm\pi^\mp}}{\B_{K\pi^\pm}}\,.
\eeq
The latter has been recently questioned~\cite{polemiqueKpi}. The problem is
that FSI effects may invalidate the notion of colour-suppression for the electroweak
penguin, thus leading to $P_{K^+\pi^-} \neq -P_{K^0\pi^+}$~\cite{neubert}. Furthermore, the 
same effects may enhance annihilation diagrams, involving a significant $V_{us}V_{ub}^\ast$ contribution to $B^+\to K^0\pi^+$ and a possibly measurable
direct CP-asymmetry in this channel~\cite{polemiqueKpi}. We will not discuss this subject
here. Rather we stress as previous authors that the $B\to K\overline{K}$ decays
may help in constraining the FSI effects~\cite{fleischerNew,KKbar}. In particular, the very
easy to detect $B\to K^+K^-$ mode is fed only by annihilations diagrams.
CLEO has already given an interesting bound on its branching ratio~\cite{CLEO}:
\beq
\B_{K^+K^-} < 0.24\times 10^{-5}\ \ \ \ \ \ \ \mbox{[90\% C.L.].}
\eeq
Thus, either the FSI effects are non-negligible and the $K^+K^-$ final state should
be detected very soon, or they are eventually out of reach of experiment and a
stringent bound on $\B_{K^+K^-}$ should be obtained~\cite{KKbar}. As claimed by the authors of 
Refs.~\cite{polemiqueKpi}, FSI effects may easily invalidate the bound~(\ref{FMbound2}); indeed, to get a significant constraint on $\gamma$, we
need the ratio $\B_{K^\pm\pi^\mp}/\B_{K\pi^\pm}$ to be sufficiently less than
1~\footnote{Note that the most recent CLEO analyses~\cite{CLEO} give $\B_{K^\pm\pi^\mp}/\B_{K\pi^\pm}\sim 1$; thus the bound~(\ref{FMbound2}) becomes
useless, even neglecting the theoretical uncertainties associated with it.} in 
order to be not too much affected by a reasonable theoretical uncertainty induced by the
neglect of electroweak penguin and annihilation contributions. On the contrary,
for the case we are interested in, namely the extraction of $\alpha$, {\it we do not
need $\B_{K^\pm\pi^\mp}/\B_{K\pi^\pm}\le 1$}, and we will see that even in the presence
of a sizeable violation of the above assumptions, we can get interesting information in the $(\rho,\eta)$ plane. In other words {\it our method concerning $\alpha$ is useful whatever the values of the branching ratios are.} However the 
Fleischer-Mannel bound is not affected by SU(3) breaking, while our method is.
Note also that Fleischer~\cite{fleischerNew} and Gronau~\cite{KKbar} have proposed very recently extensive methods which may help to
control FSI and electroweak penguin effects for the extraction of $\gamma$.

Returning to the problem of $\alpha$, we use
the above hypotheses to write $|P_{K^+\pi^-}|=|P_{K^0\pi^+}|=|A(B^+\to K^0\pi^+)|=|A(B^-\to \overline{K^0}\pi^-)|$ and thus (recall the notations~(\ref{TP}), (\ref{TP2}),
(\ref{TPK+pi-}-\ref{TPK0pi+}))
\beq\label{P=K0Pi+}
R_P\equiv |\lambda V_{cb}|^2\frac{|M^{(t)}|^2}{\B_{\pi^+\pi^-}}=\lambda^2\frac{\B_{K\pi^\pm}}{\B_{\pi^+\pi^-}}
\ \ \ \ \ 
\mbox{[assuming~{\bf  3},~{\bf  4},~{\bf  5} and~{\bf  6}].}
\eeq

The above determination of $R_P$ can be used to insert in Eq.~(\ref{masterEq6}). Of
course, the reader who does not agree with the assumptions leading to Eq.~(\ref{P=K0Pi+}) can use its own model to estimate $R_P$. Thus the method
described here is very general, and is in any case weakly model-dependent as it
depends on only one estimated parameter. The
results shown below in the $(\rho,\eta)$ plane are quite typical of what can be
obtained with such a method.

However, at this stage there is still a problem in using Eq.~(\ref{P=K0Pi+}): it is clear that we
have to give a theoretical error associated with the above determination of the
penguin amplitude. As a guess, we will simply allow a relative violation of Eq.~(\ref{P=K0Pi+}) of the order of
30\% and 60\% respectively (at the amplitude level), and leave for the future any justification of these values. Actually, {\it as long as this error is less than 100\%, the method described here
is more powerful than the bounds derived in the previous sections}
\footnote{Unfortunately, it is not clear if the relation~(\ref{P=K0Pi+}) is good
at less than 100\% relative error. Model-dependent criticisms do not predict such a huge violation of Assumption~{\bf  6}~\cite{polemiqueKpi}, however in
our case we have to take into account SU(3) breaking in Eq.~(\ref{P=K0Pi+}).}.

\begin{figure}[ht]
\centerline{\epsfxsize=0.48\textwidth\epsfbox{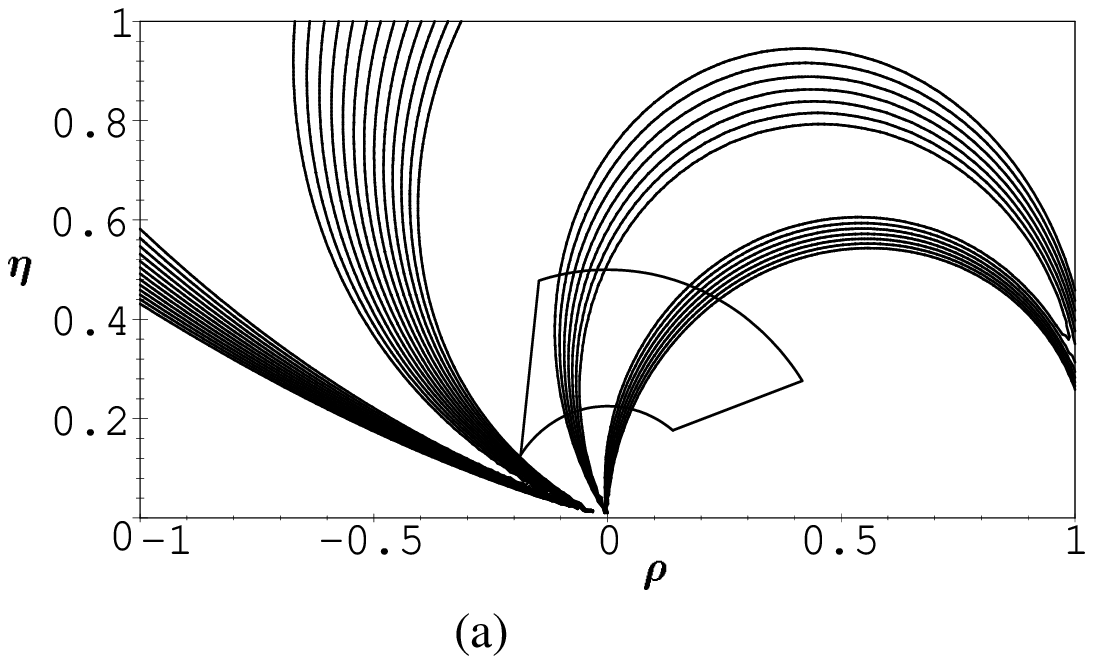}\hspace{0.01\textwidth}
\epsfxsize=0.48\textwidth\epsfbox{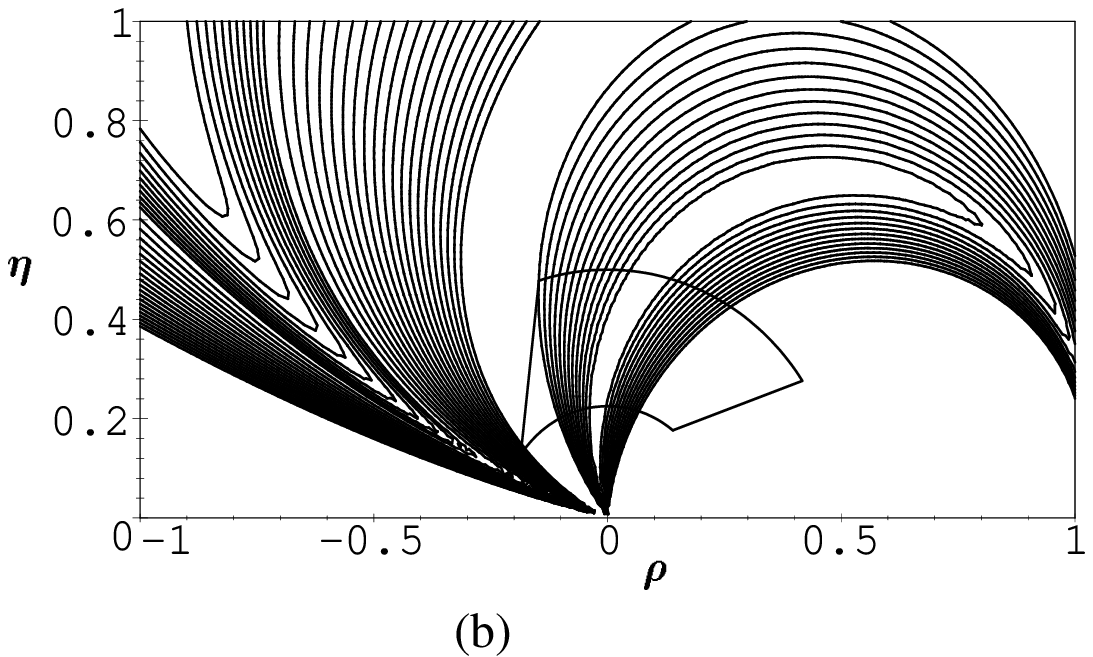}}
\caption{\it The solutions of the degree-four polynomial equation~(\ref{masterEq6}) in the
$(\rho,\eta)$ plane for the numerical example $\adir=0.12$, $\sin2\alpha\eff=0.58$ and
$R_P=\lambda^2\B_{K\pi^\pm}/\B_{\pi^+\pi^-}=0.061$ (see Appendix~\ref{model}). A guess value for the relative theoretical uncertainty on $R_P$ in Eq.~(\ref{P=K0Pi+}) is assumed: respectively 30\% {\rm(a)}
and 60\% {\rm(b)}, at the amplitude level. There
are four families of curves corresponding to the two possible signs for
$\cos2\alpha\eff$, and to the cosine discrete ambiguity of Eq.~(\ref{masterEq2}) which is hidden in the polynom~(\ref{masterEq6}). In the background is shown a crude representation of the early-1998 allowed domain~\cite{LAL2}.}
\label{fig:pronde}
\end{figure}

On Fig.~\ref{fig:pronde} we solve Eq.~(\ref{masterEq6}) with the theoretical
input~(\ref{P=K0Pi+}), and with the numerical values obtained in Appendix~\ref{model}.
Note that with our set of parameters, the Fleischer-Mannel bound becomes trivial ($\sin^2\gamma\le 1$) but it does not prevent to get useful results from Eq.~(\ref{masterEq6}). Fig.~\ref{fig:pronde} shows that with a reasonable 30\% relative violation
of the theoretical assumptions (at the amplitude level) leading to Eq.~(\ref{P=K0Pi+}), the time-dependent $B\to\pi\pi$ CP-asymmetry defines a small allowed domain in the $(\rho,\eta)$ plane, much more informative than the more
conservative bounds derived in the previous sections. This statement is quite general:
if there is a way to estimate the parameter $|P/T|$ (or $|P|$ or $R_P/R_T$ or $R_P$)
with an uncertainty of order $\sim 30\%$, then Eq.~(\ref{masterEq1}) (or Eq.~(\ref{masterEq2}) or Eq.~(\ref{masterEq5}) or Eq.~(\ref{masterEq6})) will give
rather strong constraints on $\alpha$ (or on the allowed domain in the $(\rho,\eta)$ plane). We will see in \S~\ref{GronauLondon} and~\ref{PewMasterEq} that the isospin analysis is not much
more better in this respect because it is plagued by more discrete ambiguities. Finally we stress that from the experimental point of view, our proposal is very
favourable: in addition to the usual time-dependent $B^0(t)\to\pi^+\pi^-$ CP-asymmetry,
{\it our analysis require only the measurement of the $B^\pm\to K\pi^\pm$ average branching ratio}, which is already measured (cf. Eq.~(\ref{CLEO})). In this sense
our proposal represents an improvement with respect to the Fleischer-Mannel's~\cite{FMzoology}, because the latter needs the further
knowledge of $\B_{\pi^\pm\pi^0}$ and $|V_{td}|$ (cf. \S~\ref{FMMP}).
\section{Recovering and Improving some of the Previous Approaches}
\label{previous}
In this section, we will explain how to recover in our language the Gronau-London~\cite{GL},
the Buras-Fleischer~\cite{burasK0K0b}, the Fleischer-Mannel~\cite{FMzoology} and
the Marrocchesi-Paver~\cite{paver} proposals, and in some places we will propose
improvements of these methods.
\subsection{The Gronau-London Isospin Analysis}
\label{GronauLondon}
Gronau and London have proposed a clean method to get rid of
the penguin-induced shift on $\alpha$~\cite{GL,LNQS} by measuring all the $B\to\pi\pi$ branching ratios in addition to the time-dependent CP-asymmetry~(\ref{timeCP}). Rather than repeating the
geometrical demonstration contained in the original paper, we give here the
equivalent analytical formul\ae\ and show the isospin construction in the
$\left(|P/T|,2\alpha\right)$ plane.

\begin{figure}[ht]
\centerline{\epsfxsize=0.57\textwidth\epsfbox{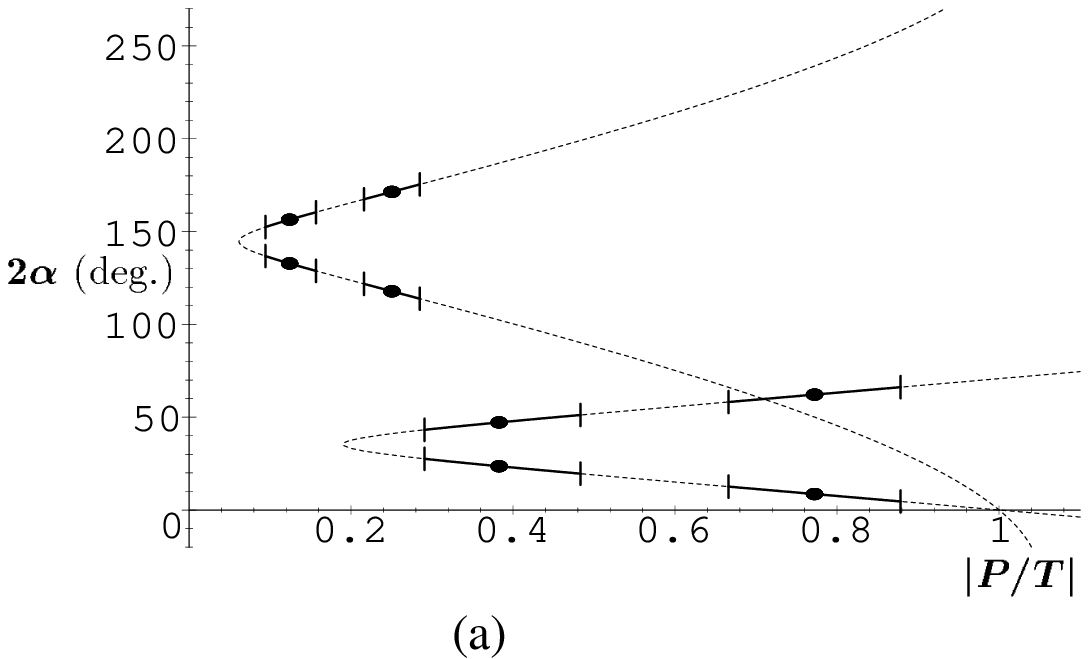}\hspace{0.01\textwidth}
\epsfxsize=0.4\textwidth\epsfbox{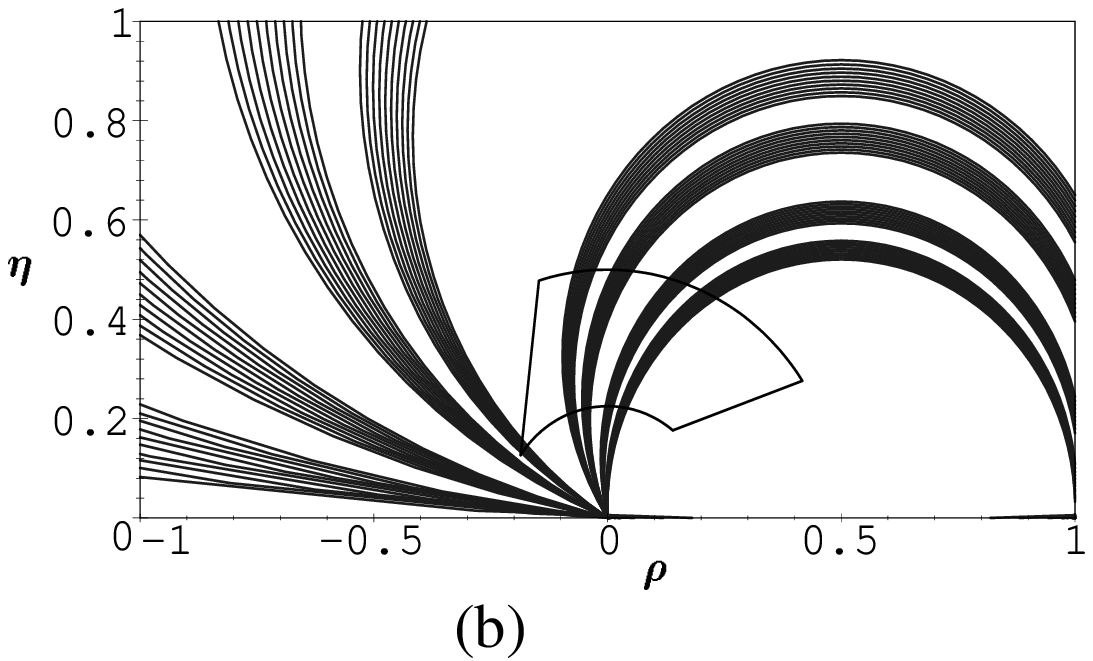}}
\caption{\it The eight solutions of the Gronau-London isospin analysis, for the numerical example $\adir=0.12$,
$\sin2\alpha\eff=0.58$,
$\B_{\pi^\pm\pi^0}/\B_{\pi^+\pi^-}=0.71$, $\B_{\pi^0\pi^0}/\B_{\pi^+\pi^-}=0.061$ and 
a direct CP-asymmetry in the $\pi^0\pi^0$ channel equal to $\adir^{\pi^0\pi^0}=0.32$ (see Appendix~\ref{model}). {\rm(a)} In the $(|P/T|,2\alpha)$ plane, the dots
represent the central values obtained from Eqs.~(\ref{phi}-\ref{phiBar-phi}),
while the solid
curves (limited by the small dashes) represent the allowed domain when assuming that
$2\alpha$ is affected by a $4^\circ$ uncertainty due to electroweak penguin
contributions, as explained in the text. {\rm(b)} The same allowed domain is represented
in the $(\rho,\eta)$ plane, where it is obtained by plotting the circle~(\ref{cercleAlpha}) for
$2\alpha$ varying in the eight solution intervals. In the background is shown a crude representation
of the early-1998 allowed domain~\cite{LAL2}.}
\label{fig:gl}
\end{figure}

The Gronau-London method 
relies on the isospin symmetry of the strong interactions: after having defined
\beqa
\Phi&\equiv&\arg \left ( A_{\pi^+\pi^0}A^\ast \right )\,,\\
\overline\Phi&\equiv&\arg \left ( \bar A_{\pi^-\pi^0}\bar A^\ast \right )\,,
\eeqa
simple trigonometry in Eqs.~(\ref{pi+pi-Isospin}-\ref{pi+pi0Isospin}) gives:
\beqa
\cos\Phi&=&\frac{1}{\sqrt{2}|A||A_{\pi^+\pi^0}|}
\left [ \frac{1}{2}|A|^2+|A_{\pi^+\pi^0}|^2-|A_{\pi^0\pi^0}|^2 \right ]\,,
\label{phi}\\
\cos\overline\Phi&=&\frac{1}{\sqrt{2}|\bar A||\bar A_{\pi^-\pi^0}|}
\left [ \frac{1}{2}|\bar A|^2+|\bar A_{\pi^-\pi^0}|^2-|\bar A_{\pi^0\pi^0}|^2 \right ]\,.\label{phiBar}
\eeqa
Eqs.~(\ref{phi}-\ref{phiBar}) are not yet sufficient to trap the penguin. However,  setting $P_{\rm EW}=0$ in~(\ref{pi+pi0Isospin}) 
implies $\arg\left ( \frac{q}{p}\bar A_{\pi^-\pi^0}A_{\pi^+\pi^0}^\ast \right ) =2\alpha$ and thus
\beq\label{phiBar-phi}
2\alpha=2\alpha\eff+\overline\Phi-\Phi\,.
\eeq
To summarize, measuring the $B\to\pi\pi$ branching ratios allows to extract the
angles $\Phi$ and $\overline\Phi$ (up to a fourfold discrete ambiguity which
corresponds to the four possible orientations of the Gronau-London triangle~\cite{GL,LNQS}) thanks to 
Eqs.~(\ref{phi}-\ref{phiBar}). As the CP-asymmetry gives $2\alpha\eff$ up to a twofold
discrete ambiguity, it is possible to get $2\alpha$ and $|P/T|$ from Eqs.~(\ref{phiBar-phi}) and~(\ref{masterEq1}) up
to an {\it eightfold} discrete ambiguity, as Fig.~\ref{fig:gl} shows.
%

\begin{figure}[ht]
\centerline{\epsfxsize=0.48\textwidth\epsfbox{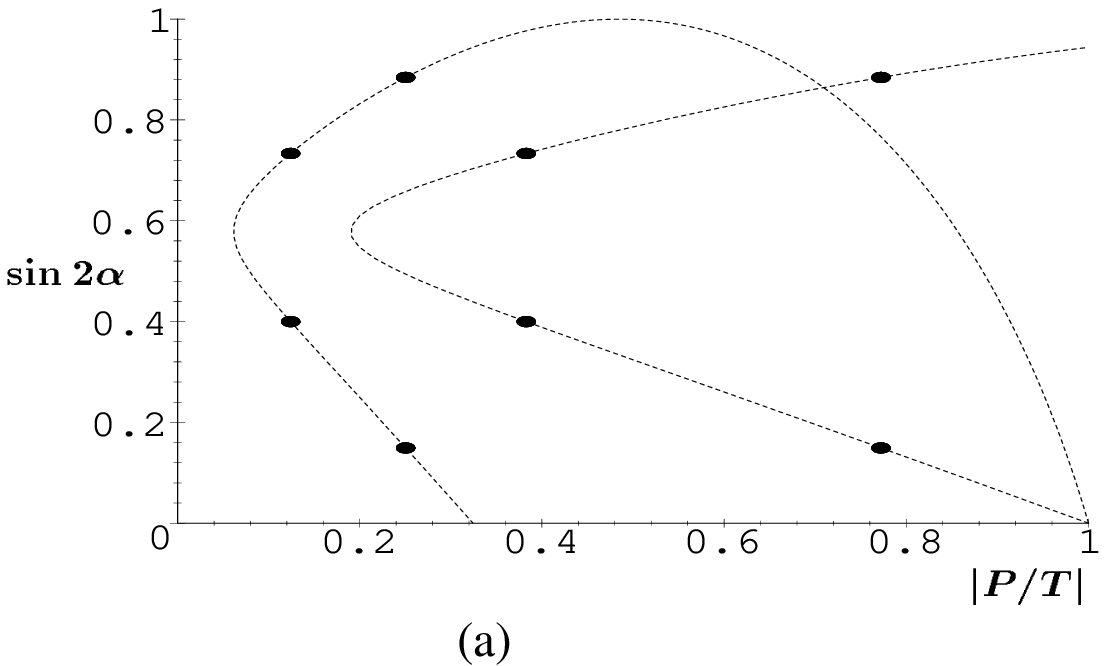}\hspace{0.01\textwidth}
\epsfxsize=0.48\textwidth\epsfbox{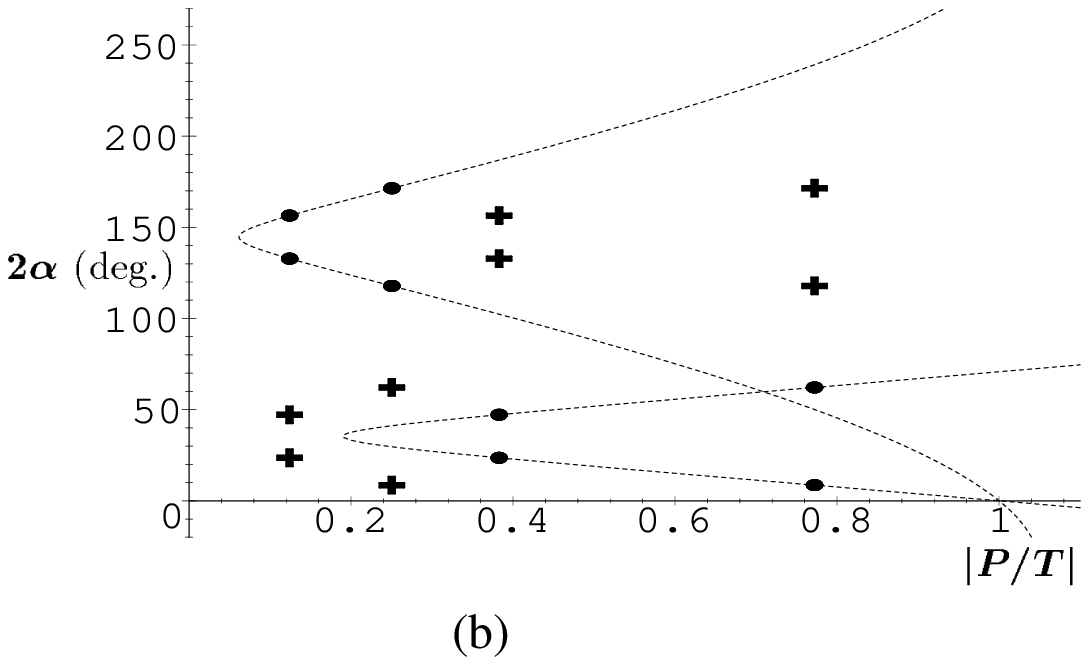}}
\caption{\it {\rm(a)} The eight solutions of the isospin
analysis in the $(|P/T|,\sin2\alpha)$ plane, for the same observables as in
Fig.~\ref{fig:gl}. {\rm(b)} The solutions in the $(|P/T|,2\alpha)$ plane obtained by computing $\arcsin(\sin2\alpha)$ and
$\pi-\arcsin(\sin2\alpha)$: the comparison with Fig.~\ref{fig:gl} shows that
the crosses are wrong solutions.}
\label{fig:glsin2a}
\end{figure}

Let us show explicitly that expressing the problem in terms of $\sin2\alpha$ is
somewhat misleading: from Fig.~\ref{fig:gl}, we can plot the eight solutions of
the isospin analysis in the $(|P/T|,\sin2\alpha)$ plane, as showed in Fig.~\ref{fig:glsin2a}(a).
Now if we forget Fig.~\ref{fig:gl} and try to get the solutions in $2\alpha$
from Fig.~\ref{fig:glsin2a}(a), we obtain the {\it sixteen} solutions of Fig.~\ref{fig:glsin2a}(b), among which eight are obviously wrong. Note two important points which have been mistreated in the original paper~\cite{GL} and to our knowledge in the subsequent literature: First, there are {\it eight}
solutions in terms of $2\alpha$ and {\it also} in terms of $\sin2\alpha$
~\footnote{If, in addition, the mixing-induced CP-asymmetry in the $\pi^0\pi^0$
is measured, there are still {\it two} solutions for $2\alpha$ (and thus four for
$\alpha$ in $[0,2\pi]$), contrary to what is said in Refs.~\cite{GL,gronauIsospin}.
In any case, the measurement of this asymmetry is expected to be very difficult.}. Second, the
isospin analysis determines $2\alpha$ rather than $\sin2\alpha$.

\subsection{Defining the Error Due to the Electroweak Penguin}
\label{PewMasterEq}
One may wonder on the size of the electroweak penguin, which is neglected in the isospin analysis.
Several authors have estimated this contribution, which turns out to be a few percents of the
dominant $B\to\pi^+\pi^-$ amplitude~\cite{Pew1,Pew2}. If this estimation is correct, then the
corresponding uncertainty on $\alpha$ may be a few degrees, which is negligible compared to
the most optimistic simulations of the statistical uncertainty~\footnote{However one should keep in mind
that the effect of the electroweak penguin on the $B\to\pi^0\pi^0$ branching ratio is not
negligible in general.}~\cite{babarBook}. In any case, a simple parametrization of the electroweak penguin 
effects can be obtained: indeed, when $P_{\rm EW}\neq 0$, there are two
new parameters, namely $\arg(P_{\rm EW}T^\ast)$ and $|P_{\rm EW}|$, and one new observable which
is the direct CP-asymmetry in the $B^\pm\to\pi^\pm\pi^0$ channel
\beq
\adir^{\pi^\pm\pi^0}=\frac{\BR(B^+\to\pi^+\pi^0)-\BR(B^-\to\pi^-\pi^0)}{\BR(B^+\to\pi^+\pi^0)+\BR(B^-\to\pi^-\pi^0)}
\eeq
which vanishes when $P_{\rm EW}\to 0$.
Similarly to the case of
the strong penguin, as discussed at length in this paper, it is possible to express $\alpha$
as a simple function of the observables of the Gronau-London isospin analysis, the direct CP-asymmetry in the $B^\pm\to\pi^\pm\pi^0$ channel and the unknown parameter $|P_{\rm EW}|$. The same technique leading to Eq.~(\ref{masterEq1}) allows to find
\beq\label{masterEq8}
\cos(2\alpha-2\alpha^\prime\eff)
=\frac{1}{\sqrt{1-\left[\adir^{\pi^\pm\pi^0}\right]^2}}
\left[1-\left(1-\sqrt{1-\left[\adir^{\pi^\pm\pi^0}\right]^2}
\cos2\alpha^\prime\eff\right)
\left|\frac{P_{\rm EW}}{T_{\pi^+\pi^0}}\right|^2\right]
\eeq
where $2\alpha^\prime\eff$ is the value of $2\alpha$ when $P_{\rm EW}=0$ (see Eq.~(\ref{phiBar-phi}))
\beq\label{a'eff}
2\alpha^\prime\eff=2\alpha\eff+\overline{\Phi}-\Phi\,.
\eeq

Thus Eqs.~(\ref{masterEq8}-\ref{a'eff}) describe the departure from the isospin analysis~(\ref{phiBar-phi}) due to the electroweak penguin contributions. As a particular case, we obtain the bound
\beq\label{PewBound}
|2\alpha-2\alpha^\prime\eff|\le\arccos\left[1-2\left|\frac{P_{\rm EW}}{T_{\pi^+\pi^0}}\right|^2\right]\sim2\left|\frac{P_{\rm EW}}{T_{\pi^+\pi^0}}\right|
\eeq
which was derived in Ref.~\cite{Pew1}. Note that the ratio $|P_{\rm EW}/T_{\pi^+\pi^0}|$ is rather independent of the size of the colour-suppression,
although the impact of $|P_{\rm EW}|$ on BR$(B\to\pi^0\pi^0)$ is not negligible.

Thus, whatever the way to estimate the parameter
$|P_{\rm EW}/T_{\pi^+\pi^0}|$, one is lead to a simple
and weakly model-dependent definition of the theoretical error on $\alpha$ induced by the electroweak penguin. Using factorization for the estimation of the r.h.s.
of~(\ref{PewBound}), we find typically $|2\alpha-2\alpha^\prime\eff|\lsim 4^\circ$;
this error has been reported on Fig.~\ref{fig:gl} for illustration. This figure
shows the sensitivity of the isospin analysis with respect to the discrete
ambiguities: with an error on $\alpha$ as small as $2^\circ$ (here this error
comes from the electroweak penguin contributions, but unfortunately there are also the uncertainties of experimental origin which we have not considered), the
four separate solutions (for a given sign of $\cos2\alpha\eff$) tends to merge
quite quickly. This is not {\it a posteriori} surprising: indeed these four
solutions are separated because of the QCD penguin contributions, i.e. because
of a relatively small effect; they become degenerate in the no-penguin limit.
Comparison with Figs.~\ref{fig:borne} and~\ref{fig:pronde} suggests actually that
the more simpler approaches to control the penguin effects described in the previous sections may be competitive with the more complete isospin analysis, unless the
observables of the latter are known with a very high accuracy.

The main drawback of the Gronau-London analysis is the expected rarity of the
$B\to\pi^0\pi^0$ channel, which branching ratio is expected to be about $10^{-7}$--$10^{-6}$.
The neutral pions are not easy to detect, and one needs to tag the flavour of the
$B$-meson in order to get separately $|A_{\pi^0\pi^0}|$ and $|\bar A_{\pi^0\pi^0}|$,
according to Eqs.~(\ref{phi}-\ref{phiBar}). The small number of effectively useful
events expected at an $e^+e^-$ $B$-factory constitutes a difficult challenge to
the experimentalists while the impossibility to detect two neutral pions in future
hadronic machines does not improve the situation. This shows the interest of the bounds~(\ref{GQbound}) and~(\ref{pi0pi0Bound}).

\subsection{The Buras-Fleischer Proposal}
\label{BF}
Considering the experimental difficulties associated with the Gronau-London
analysis, Buras and Fleischer have proposed an alternative way to get rid of the penguin uncertainty, using SU(3) and the time-dependent CP-asymmetry of the pure penguin mode $B\to K^0\overline{K^0}$ ~\cite{burasK0K0b}. They argue that the SU(3) breaking effects are of the
same order as the electroweak penguin uncertainty of the isospin analysis.

The idea is simple: similarly to the $B\to\pi^+\pi^-$ decay we define the time-dependent
$B\to K^0\overline{K^0}$ CP-asymmetry
\beq
a^{K\bar K}_{\rm CP}(t)=\adir^{K\bar K}\cos\Delta m\,t-\sqrt{1-\left[\adir^{K\bar K}\,\right]^2}\sin2\alpha\eff^{K\bar K} \sin\Delta m\,t \,,
\eeq
where we have used the notation $\sin2\alpha\eff^{K\bar K}$ to make apparent the
resemblance with Eq.~(\ref{timeCP}): we stress however that $2\alpha\eff^{K\bar K}$
reduces to $2\alpha$ when the $T_{K\overline{K}}$ amplitude dominates in Eq.~(\ref{TPK0K0b}), i.e. when the difference between the $u$- and $c$- penguins
dominates over the difference between the $t$- and $c$- penguins, which is presumably
an extreme case. Conversely, in the absence of long-distance $u$- and $c$- penguins, we have $\sin2\alpha\eff^{K\bar K}=\adir^{K\bar K}=0$~\cite{fleischerK0K0b,burasK0K0b}.

Thus, following Eq.~(\ref{masterEq2}) we find
\beq\label{PKKb}
|P_{K^0\overline{K^0}}|^2=\frac{\B_{K^0\overline{K^0}}}{1-\cos2\alpha}\left [ 1-\sqrt{1-\left[\adir^{K\bar K}\,\right]^2}\cos(2\alpha-2\alpha\eff^{K\bar K})\right ]\,,
\eeq
which reduces to $|P_{K^0\overline{K^0}}|^2=\B_{K^0\overline{K^0}}$ if the top-penguin dominates the amplitude.

\begin{figure}[ht]
\centerline{\epsfxsize=0.57\textwidth\epsfbox{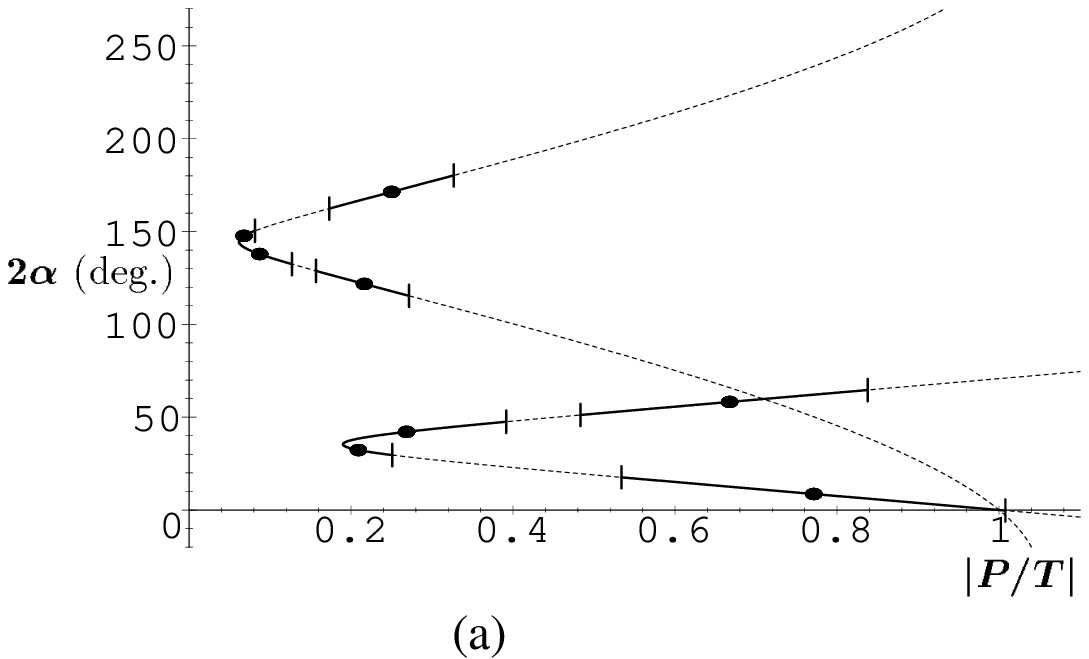}\hspace{0.01\textwidth}
\epsfxsize=0.4\textwidth\epsfbox{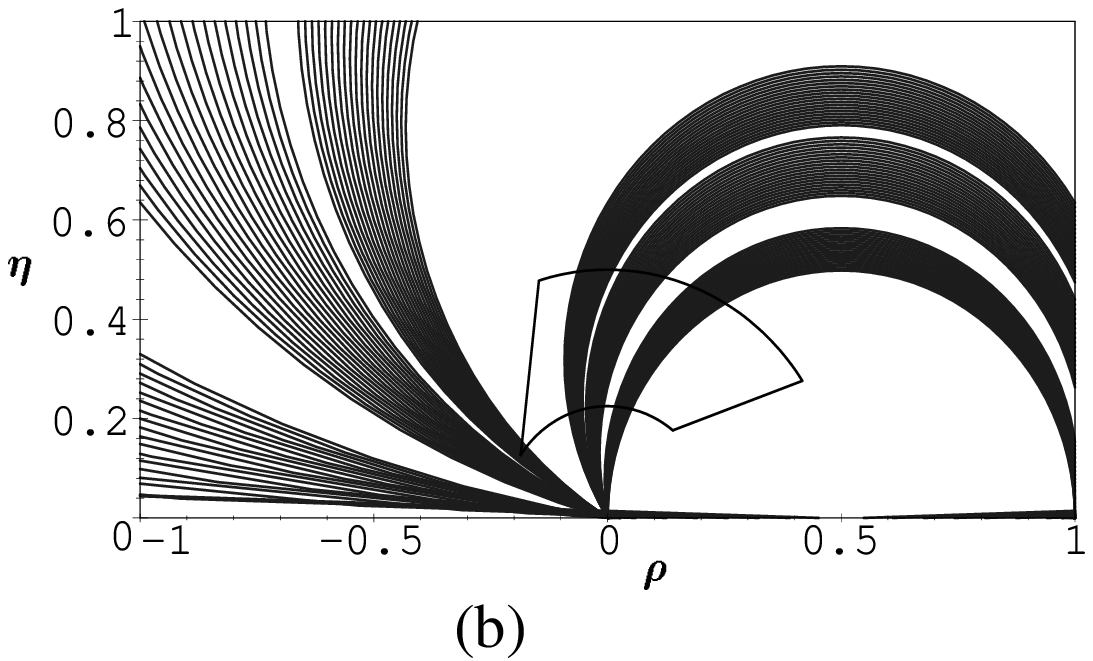}}
\caption{\it The eight solutions of the Buras-Fleischer analysis, for the numerical example $\adir=0.12$,
$\sin2\alpha\eff=0.58$,
$\adir^{K\bar K}=0.21$, $\sin2\alpha\eff^{K\bar K}=0.059$ and $\B_{K^0\overline{K^0}}/\B_{\pi^+\pi^-}=0.058$ (see Appendix~\ref{model}). {\rm(a)} In the $(|P/T|,2\alpha)$ plane, the dots represent the central values obtained from
Eqs.~(\ref{Dronde}-\ref{2a-psi}), while the solid curves (limited by the small
dashes) represent the allowed domain when assuming that Eq.~(\ref{P=PKKb}) is
affected by a guess 30\% relative uncertainty. Some of the solutions merge
because of this theoretical error, leaving only six separate solutions.
{\rm(b)} The same allowed domain is represented in the $(\rho,\eta)$
plane, where it is obtained by plotting the circle~(\ref{cercleAlpha}) for $2\alpha$ varying in
the six solution intervals. In the background
is shown a crude representation of the early-1998 allowed domain~\cite{LAL2}.}
\label{fig:bf}
\end{figure}

Assuming SU(3) and neglecting the (colour-suppressed) electroweak penguin contributions, we may write
\beq\label{P=PKKb}
|P|=|P_{K^0\overline{K^0}}|\,.
\eeq
Thus from Eqs.~(\ref{masterEq2}),~(\ref{PKKb}) and~(\ref{P=PKKb}) we have
\beq\label{zozo}
\left [ 1-\sqrt{1-\adir^2}\cos(2\alpha-2\alpha\eff)\right]
-\frac{\B_{K^0\overline{K^0}}}{\B_{\pi^+\pi^-}}\left [ 1-\sqrt{1-\left[\adir^{K\bar K}\,\right]^2}\cos(2\alpha-2\alpha\eff^{K\bar K})\right ]=0\,.
\eeq
Defining the following quantity that can be written in terms of observables
\beqa
{\cal D}&=&\sqrt{1-\adir^2}\,\exp(i\,2\alpha\eff)-\frac{\B_{K^0\overline{K^0}}}{\B_{\pi^+\pi^-}}
\sqrt{1-\left[\adir^{K\bar K}\,\right]^2}\,\exp(i\,2\alpha\eff^{K\bar K}) \nn\\
&\equiv& |{\cal D}|e^{i\Psi} \label{Dronde}\,,
\eeqa
Eq.~(\ref{zozo}) becomes
\beq\label{2a-psi}
\cos(2\alpha-\Psi)=\frac{1}{|{\cal D}|}\left[1-\frac{\B_{K^0\overline{K^0}}}{\B_{\pi^+\pi^-}}\right] \,.
\eeq
As $2\alpha\eff$ and $2\alpha\eff^{K\bar K}$ are both measured up to a twofold discrete ambiguity, Eq.~(\ref{2a-psi}) gives $2\alpha$ up to an eightfold discrete
ambiguity. An explicit example is given on Fig.~\ref{fig:bf}.

However from the experimental point of view
the study of this decay may be as difficult as the isospin analysis: First, it is a pure
$b\to d$ penguin decay and is thus expected to be very rare ($\sim 10^{-7}$--$10^{-6}$)~\cite{rome}. And second, the time-dependence of the decay rate may
be difficult to reconstruct because the neutral kaons decay far away from the primary vertex.
This shows the interest of the bound~(\ref{K0K0bBound}), very symmetrically to the case of the isospin analysis.

\subsection{The Fleischer-Mannel and Marrocchesi-Paver Methods}
\label{FMMP}
Fleischer and Mannel~\cite{FMzoology}, as well as Marrocchesi and Paver~\cite{paver}
had already remarked that knowing the value of $|P/T|$ alone leads to the extraction of $\alpha$. Therefore they have used Eq.~(\ref{masterEq1}) without explicitly
having written it, and without having noticed the complete generality of the method. Let us briefly sketch the main points of their studies:
\begin{itemize}
\item
Fleischer and Mannel use a first-order expansion in $|P/T|$. We have shown that this
approximation, although numerically good, is unnecessary: Eq.~(\ref{masterEq1}) is 
exact and not more complicated than its first-order expansion.
\item
Fleischer and Mannel estimate $|P/T|$ by assuming Eq.~(\ref{P=K0Pi+}) and neglecting
the colour-suppressed contributions to $B^\pm\to\pi^\pm\pi^0$~\cite{FMzoology}
\beq\label{FMPsurT}
\left|\frac{P}{T}\right|^2=\left|\frac{V_{td}V_{tb}^\ast}{V_{cs}V_{cb}^\ast}\right|^2\times
\frac{\B_{K\pi^\pm}}{2\B_{\pi^\pm\pi^0}} \,,
\eeq
while Marrocchesi and Paver use factorization to calculate (in this case $|P/T|$
is just proportional to a ratio of short-distance Wilson coefficients times a CKM factor)~\cite{paver}
\beq\label{PMPsurT}
\left|\frac{P}{T}\right|=\frac{\sin(\alpha+\beta)}{\sin\beta}\times0.055 \,.
\eeq
The two above equations represent alternatives to the method presented in
Section~\ref{K0pi+Section}, although in the second case it is not clear to what extent factorization
can be used to calculate $|P/T|$~\cite{rome}. Note that these two approaches
use a single model-dependent input, as the method we have proposed in Section~(\ref{K0pi+Section}).
\item
Both Fleischer and Mannel and Marrocchesi and Paver face the problem of knowing
$|V_{td}|$ or $\sin(\alpha+\beta)/\sin\beta$. The first two authors assume simply
that $|V_{td}/(\lambda V_{cb})|$ is known from CP-conserving measurements~\cite{FMzoology}, while the second two authors take the value of
$\beta$ as it would be given by the future measurements of the $B\to J/\Psi K_S$ CP-asymmetry and obtain an equation depending on $\alpha$ alone~\cite{paver}. However it is not clear if CP-conserving measurements will give
$|V_{td}/(\lambda V_{cb})|$ with enough accuracy, and using instead the value of
$\beta$ unfortunately propagates the uncertainty and the discrete ambiguities associated
with the measurement of $\beta$ into the extraction of $\alpha$. We have shown in 
\S~\ref{RhoEtaPlane} that one can avoid these problems by directly
writing easy-to-solve polynomial equations in the $(\rho,\eta)$ plane, therefore
without invoking other independent CKM measurements. For the Fleischer-Mannel
proposal one should write $|P/T|^2=\lambda^2|1-\rho-i\eta|^2\times\B_{K\pi^\pm}/(2\B_{\pi^\pm\pi^0})$ and report this expression into Eq.~(\ref{masterEq1}) to obtain an equation~\footnote{We have not written this equation, which is not Eq.~(\ref{masterEq5}), because the $\B_{K\pi^\pm}/(2\B_{\pi^\pm\pi^0})$ ratio already incorporates a $|V_{ub}^\ast|$ factor.} in the
variables $(\rho,\eta)$, without the need to know $|V_{td}/(\lambda V_{cb})|$. For the Marrocchesi-Paver method
one should simply insert $R_P/R_T=0.055$ in Eq.~(\ref{masterEq5}) independently of $\beta$. Thus our
framework allows to improve significantly these proposals.
\item
Finally we would like to stress once again the importance of the discrete ambiguities.
While they are not discussed at all by Fleischer and Mannel~\cite{FMzoology}, we 
believe that the treatment of Marrocchesi and Paver is incomplete: for a given value
of $|P/T|$ (inferred from factorization and a given value for $\beta$), they find
two solutions for $\alpha$ between 0 and $\pi$. We have shown in \S~\ref{masterSection}
that there are four such solutions which, because of the finiteness of the errors
(both theoretical and experimental),
may merge among themselves.
\end{itemize}
\section{Conclusion}
We have shown that in the presence of penguin contributions, the information on the
CKM angle $\alpha$ coming from the measurement of the time-dependent $B^0(t)\to\pi^+\pi^-$ CP-asymmetry can be summarized in a set of simple equations, expressing $\alpha$ as a multi-valued function of a single theoretically unknown
parameter. These equations, free of any assumption besides the Standard Model, provide by themselves an exact model-independent interpretation of future CP-experiments.

It is also possible to choose as the unknown a pure QCD quantity, in which case the
above equations should be expressed directly in the $(\rho,\eta)$ plane, thanks to
the unitarity of the CKM matrix which predicts relations between the CP-violating angles and the CP-conserving sides of the Unitarity Triangle. Whatever the choice
of the single unknown, as for example the ratio of penguin to tree matrix elements, this unavoidable non-perturbative parameter in $B\to\pi^+\pi^-$ could be compared to $B_K$ in the kaon system which allows to report the measurement of $\epsilon_K$ in the $(\rho,\eta)$ plane. However the ratio $|P/T|$ is a much more complicated quantity than $B_K$, and would be very difficult to obtain from QCD fundamental methods.

Using these analytic expressions, we have assumed some reasonable hypotheses to constrain the free parameter. Doing so we have derived several new bounds on the
penguin\--induced shift $|2\alpha-2\alpha\eff|$, generalizing the result of Grossman
and Quinn~\cite{GQbound}. One of these bounds is determined by the ratio
$\lambda^2\B_{K^\pm\pi^\mp}/\B_{\pi^+\pi^-}$ on which one would have an experimental
value very soon.

Accepting less conservative assumptions, stronger constraints on $\alpha$ can be
obtained. For example in the limit where the annihilation and electroweak penguin diagrams can be neglected,
and using SU(3), the knowledge of the $B^\pm\to K\pi^\pm$ branching ratio is a
sufficient information to extract the theoretical unknown. Assuming a reasonable
30\% relative uncertainty (at the amplitude level) on the unavoidable hypotheses,
a relatively small allowed domain in the $(\rho,\eta)$ plane can be found, independently
of any other measurement. This
method could be competitive with the full Gronau-London isospin analysis, because the
latter is plagued by twice more discrete ambiguities. From the experimental point
of view, our proposal may be much more easier to achieve. More generally, if by some
other argument a knowledge of the modulus of the penguin amplitude---or the ratio
of penguin to tree---with a
$\sim 30\%$ uncertainty can be achieved, then rather strong constraints on $\alpha$
should be obtained.

However we do not pretend that the theoretical uncertainty on $\alpha$ will be small.
Rather we believe that this error may be quite well controlled by conservative arguments.
This shows the importance of generalizing our framework to other channels sensitive
to $\alpha$: if we are unlucky in the $\pi^+\pi^-$ channel, it may happen that
we are lucky in others. As the problem of the discrete ambiguities is crucial in these analyses,
the modes providing new CP-observables are of a particular interest: for example measuring directly the sign of $\cos2\alpha\eff$ rather than determining it from the SM constraints
on the UT would be a valuable information, even in the presence of sizeable penguin contributions, as it would allow to reduce the discrete ambiguities
generated when expressing $\alpha$ as a function of the observables and of one
model-dependent input. It has been shown previously~\cite{snyderQuinn} that the analysis of the $B\to\rho\pi\to 3\pi$ Dalitz plot actually leads to the measurement
of a kind of $\cos2\alpha\eff$~\footnote{Eventually the $B\to\rho\pi\to 3\pi$ time-dependent Dalitz plot together
with the isospin symmetry also allows the extraction of penguins~\cite{snyderQuinn}.
However, such an analysis seems to require a high statistics~\cite{babar3pi}.} (which is of course different from the $\alpha\eff$
in $B\to\pi\pi$), and we are currently studying the possibility to describe this
interesting decay similarly to $B\to\pi\pi$~\cite{babar3pi}. Likewise the angular distribution of the decay $B\to\Lambda\overline{\Lambda}$ contains also terms proportional to the cosine
of an effective $\alpha$ angle~\cite{hyperons}.

It is quite clear that all the strategies proposed until now to disentangle the penguin 
pollution in various channels will give different information on $\alpha$, each
relying on very different theoretical assumptions and on different observables.
Our framework allows to treat all these sources of information in a transparent
and unified way. Thus we will have certainly a strong cross-check of the various
results. If this cross-check is successful, we may think to combine these results
in order to have a more precise knowledge of $\alpha$. However, we are aware that combining theoretical and
experimental errors is a difficult problem by itself which is beyond the scope
of the present paper.
\section*{Acknowledgements}
I acknowledge Y. Grossman, A. Jacholkowska, F. Le Diberder, G. Martinelli, T. Nakada, S. Plaszczynski, M.-H. Schune, L. Silvestrini and S. Versill\'e for useful
discussions and comments. I am also grateful to L. Bourhis for help. Finally I am indebted to A. Le Yaouanc, L. Oliver, O. P\`ene and
J.-C. Raynal, without whom this work could not have been achieved, for constant and stimulating encouragements and for a careful reading of
the manuscript.

\appendix
\section{A Typical Set of Theoretical Parameters}
\label{model}
In this appendix, we define a typical set of parameters in order to compute the
relevant observables. We assume that~{\bf 1--6} are
exact, and neglect further {\it all} annihilation diagrams. Thus the amplitudes in
Eqs.~(\ref{TP}-\ref{TPpi+pi0}), (\ref{TPK0K0b}-\ref{TPK0pi+}) write:
\beqa
A(B^0\to\pi^+\pi^-)&=&e^{i\gamma}\,T+e^{-i\beta}\,P\,,\\
A(B^0\to\pi^0\pi^0)&=&\frac{1}{\sqrt{2}}\left[e^{i\gamma}\left(T-T_{\pi^+\pi^0}\right)-e^{-i\beta}\,P\right]\,,\\
A(B^+\to\pi^+\pi^0)&=&\frac{1}{\sqrt{2}}e^{i\gamma}\,T_{\pi^+\pi^0}\,,\\
A(B^0\to K^0\overline{K^0})&=&P\left(e^{-i\beta}+\left|\frac{V_{ub}^\ast}{V_{td}}\right|e^{i\gamma}\,r_u\right)\,,\\
A(B^0\to K^+\pi^-)&=&\lambda e^{i\gamma}\,T+\left|\frac{V_{ts}}{\lambda V_{td}}\right|e^{i(\delta^\prime-\delta)}P\,,\\
A(B^+\to K^0\pi^+)&=&\left|\frac{V_{ts}}{\lambda V_{td}}\right|e^{i\delta^\prime}P\,.
\eeqa
Note that in the strict SU(3) limit and neglecting annihilation diagrams, $\delta^\prime=\delta$.

\begin{table}[ht]\begin{center}
\begin{tabular}{c|c|c|c|c|c}
$\B_{\pi^+\pi^-}$ (normalization) & $\B_{\pi^0\pi^0}$ & $\B_{\pi^\pm\pi^0}$ &
$\B_{K^0\overline{K^0}}$ & $\B_{K^\pm\pi^\mp}$ & $\B_{K\pi^\pm}$ \\\hline
0.75 & 0.0455 & 0.533 & 0.0433 & 1.075 & 0.948
\end{tabular}
\caption{\it The various average branching ratios (in units of $10^{-5}$) from our set of parameters. They are consistent with CLEO data~(\ref{CLEO}).}\label{tableBR}
\end{center}\end{table}
\begin{table}[ht]\begin{center}
\begin{tabular}{c|c|c|c|c|c}
$\adir$ & $\sin2\alpha\eff$ & $\adir^{\pi^0\pi^0}$ &
$\adir^{K^0\overline{K^0}}$ & $\sin2\alpha\eff^{K^0\overline{K^0}}$ &
$\adir^{K^\pm\pi^\mp}$ \\\hline
0.117 & 0.579 & 0.317 & 0.209 & 0.0592 & 0.108
\end{tabular}
\caption{\it The various CP-asymetries from our set of parameters.}\label{tableCP}
\end{center}\end{table}

Numerically, we take $\B_{\pi^+\pi^-}=0.75\times 10^{-5}$, which fixes the normalization
of the amplitudes and choose $T$ real which fixes the origin of phases. Then we choose $|P/T|=0.25$ which is a quite sizeable value (cf. \S~\ref{magnitude}) and $\delta=-15^\circ$ which is a large violation of naive
factorization which gives $\delta=180^\circ$. The normalization is then given by
$|T|=0.826\times\sqrt{10^5\B_{\pi^+\pi^-}}$ (in ``units of two-body branching ratio''). We choose also $T_{\pi^+\pi^0}=1.25e^{-i7^\circ}\times|T|$ which takes into account the usual $a_2\sim0.25$ colour-suppression factor and some FSI phases,
$\delta^\prime=+20^\circ$, and $r_u=0.3e^{+i75^\circ}$ which is a ratio of long-distance over short-distance penguin matrix elements. For the CKM parameters,
we have $\lambda=0.2205$ and take $\rho=0.10$, $\eta=0.34$ which is around
the center of the early-1998 allowed domain ($\alpha=85.7^\circ$)~\cite{LAL}. The resulting values for
the observables are summarized in Tables~\ref{tableBR} and ~\ref{tableCP}.
Let us stress that these
values are only indicative and that the real numbers may be very different. Our
set of parameters results from a compromise between the need to take into account
various effects in a more or less realistic way and the pedagogical needs (for
example it is easier to discuss the number of discrete solutions when they are
quite well separated, which is often not the case in practice).
Finally we notice that the penguin-induced shift on $\alpha$ is quite large for this set
of parameters: $2\alpha-2\alpha\eff=+26.7^\circ$.
\section{Bounds Independent of Direct CP-Violation}
\label{2aeffbar}
Here our purpose
 is to derive bounds which are fully independent of $\adir$, and
thus are not affected by the experimental uncertainty associated with the measurement
of direct CP-violation~\cite{jeanclaude}. As
far as the bound~(\ref{GQbound2}) is concerned, a different demonstration has been given
by Grossman and Quinn~\cite{GQbound}.

Consider the bounds~(\ref{GQbound2}-\ref{bound5}): they all can be written
\beq\label{E1}
\frac{1-m}{\sqrt{1-\adir^2}} \,\le\, \cos(2\alpha-2\alpha\eff)
\eeq
where $m$ is a positive ratio of branching ratios and is expected to be smaller than 2 (otherwise the
bound is useless).
If $\adir$ is not known, then $2\alpha\eff$ is not known either. Rather one gets
from the $\sin\Delta m\,t$ term in~(\ref{timeCP2}) the effective angle
\beq\label{E2}
\sin2\overline{\alpha}\eff\equiv\sqrt{1-\adir^2}\sin2\alpha\eff\,.
\eeq
Since $|\sin2\overline{\alpha}\eff|\le|\sin2\alpha\eff|$ one has:
\beq\label{E3}
|\cos2\alpha\eff| \,\le\, |\cos2\overline{\alpha}\eff|\,.
\eeq
As the sign of $\cos2\overline{\alpha}\eff$ is not observable, it can be chosen
arbitrarily. It is convenient to define
\beq
{\rm sign}(\cos2\overline{\alpha}\eff)\equiv{\rm sign}(\cos2\alpha)\,,
\eeq
in such a way that~(\ref{E3}) gives
\beq\label{E4}
|\cos2\alpha\eff\cos2\alpha| \,\le\, |\cos2\overline{\alpha}\eff\cos2\alpha|=\cos2\overline{\alpha}\eff\cos2\alpha\,.
\eeq
Thus Eqs.~(\ref{E1}) and~(\ref{E4}) imply
\beqa
1-m &\le& \sqrt{1-\adir^2}\cos2\alpha\eff\cos2\alpha+\sin2\overline{\alpha}\eff\sin2\alpha\nn\\
&\le&|\cos2\alpha\eff\cos2\alpha|+\sin2\overline{\alpha}\eff\sin2\alpha\nn\\
&\le&\cos2\overline{\alpha}\eff\cos2\alpha+\sin2\overline{\alpha}\eff\sin2\alpha=\cos(2\alpha-2\overline{\alpha}\eff)\,,
\eeqa
and we obtain the announced result, namely 
\beq
\frac{1-m}{\sqrt{1-\adir^2}} \,\le\, \cos(2\alpha-2\alpha\eff)
\ \Rightarrow\ 1-m \,\le\, \cos(2\alpha-2\overline{\alpha}\eff)\,.
\eeq

It is straightforward to demonstrate an analogous result for the bound~(\ref{bound1}).


\end{document}